\documentclass[prodmode,acmtois]{acmsmall} 

\usepackage[ruled]{algorithm2e}
\usepackage{subfigure}
\usepackage{amsmath}
\usepackage{epstopdf}

\newcommand{\paratitle}[1]{\vspace{1.5ex}\noindent\textbf{#1}}
\newcommand{\ie}{\emph{i.e.,}\xspace}
\newcommand{\eg}{\emph{e.g.,}\xspace}

\SetAlFnt{\small}
\SetAlCapFnt{\small}
\SetAlCapNameFnt{\small}
\SetAlCapHSkip{0pt}
\IncMargin{-\parindent}

\acmVolume{35}
\acmNumber{4}
\acmArticle{8}
\acmYear{2017}
\acmMonth{10}

\doi{0000001.0000001}

\issn{1234-56789}

\begin{document}

\markboth{..}{A Neural Network Approach to Joint Modeling Social Networks and Mobile Trajectories }

\title{A Neural Network Approach to Joint Modeling Social Networks and Mobile Trajectories}
\author{Cheng Yang$^\dag$ 
\affil{Tsinghua University, China}
Maosong Sun
\affil{Tsinghua University, China}
Wayne Xin Zhao$^{*}$
\affil{Renmin University of China}
Zhiyuan Liu$^{*}$
\affil{Tsinghua University, China}
Edward Y.Chang
\affil{HTC Research \& Innovation}
}

\begin{abstract}
The accelerated growth of mobile trajectories in location-based services brings valuable data resources to understand users' moving behaviors.
Apart from recording the trajectory data, another major characteristic of these location-based services is that they also allow the users to connect whomever they like or are interested in. A combination of social networking and location-based services is called as location-based social networks (LBSN).
As shown in \cite{cho2013socially}, locations that are frequently visited by socially-related persons tend to be correlated, which indicates the close association between social connections and trajectory behaviors of users in LBSNs. In order to better analyze and mine LBSN data, we need to have a comprehensive view to analyze and mine the information from the two aspects, \ie the social network and mobile trajectory data.

Specifically, we present a novel neural network model which can jointly model both social networks and mobile trajectories. Our model consists of two components: the construction of social networks and the generation of mobile trajectories. First we adopt a network embedding method for the construction of social networks: a networking representation can be derived for a user. The key of our model lies in the component of generating mobile trajectories. Secondly, we consider four factors that influence the generation process of mobile trajectories, namely user visit preference, influence of friends, short-term sequential contexts and long-term sequential contexts. To characterize the last two contexts, we employ the RNN and GRU models to capture the sequential relatedness in mobile trajectories at different levels, i.e., short term or long term. Finally, the two components are tied by sharing the user network representations. Experimental results on two important applications demonstrate the effectiveness of our model. Especially, the improvement over baselines is more significant when either network structure or trajectory data is sparse.

\end{abstract}

%
%


%
%

\terms{Algorithms, Performance}


\acmformat{..}

\begin{bottomstuff}
$^\dag$ This work was done during the first author's internship at HTC Beijing Research.

$^{*}$ Corresponding authors

This work was supported by the 973 Program (No. 2014CB340501), the Major Project of the National Social Science Foundation of China (13\&ZD190), the National Natural Science Foundation of China (NSFC No.61502502, 61572273 and 61532010), Tsinghua University Initiative Scientific Research Program (20151080406) and Beijing Natural Science Foundation under the grant number 4162032.

Authors¡¯ addresses: Cheng Yang, Maosong Sun and Zhiyuan Liu, Department of Computer Science and Technology, Tsinghua University, Beijing 100084; emails: cheng-ya14@mails.tsinghua.edu.cn, sms@mail.tsinghua.edu.cn, liuzy@tsinghua.edu.cn; W.X.Zhao, School of Information \& Beijing Key Laboratory of Big Data Management and Analysis Methods, Renmin University of China, Beijing 100872; email: batmanfly@gmail.com; E.Y.Chang, HTC Research \& Innovation, Palo Alto, CA 94306; email: eyuchang@gmail.com.
\end{bottomstuff}

\maketitle

\section{Introduction}
In recent years,  mobile devices (\eg smartphones and tablets) are widely used almost everywhere. With the innovation and development on Internet technology,
mobile devices have become an essential connection to the broader world of online information for users. In daily life, a user can utilize her smartphone for
 conducting many life activities, including researching a travel plan, accessing online education, and looking for a job. The accelerated growth of mobile usage brings a unique opportunity to data mining research communities. Among these rich mobile data, an important kind of data resource is the huge amount of mobile trajectory data obtained from GPS sensors on mobile devices.  These sensor footprints provide a valuable
information resource to discover users' trajectory patterns and understand their moving behaviors. Several location-based sharing services have emerged and received much attention, such as \emph{Gowalla}\footnote{https://en.wikipedia.org/wiki/Gowalla} and \emph{Brightkite}\footnote{https://en.wikipedia.org/wiki/Brightkite}.

\begin{figure}[htb]
\centering
\begin{minipage}{\columnwidth}
\subfigure[Friendship Network]{\includegraphics[width=0.45\columnwidth]{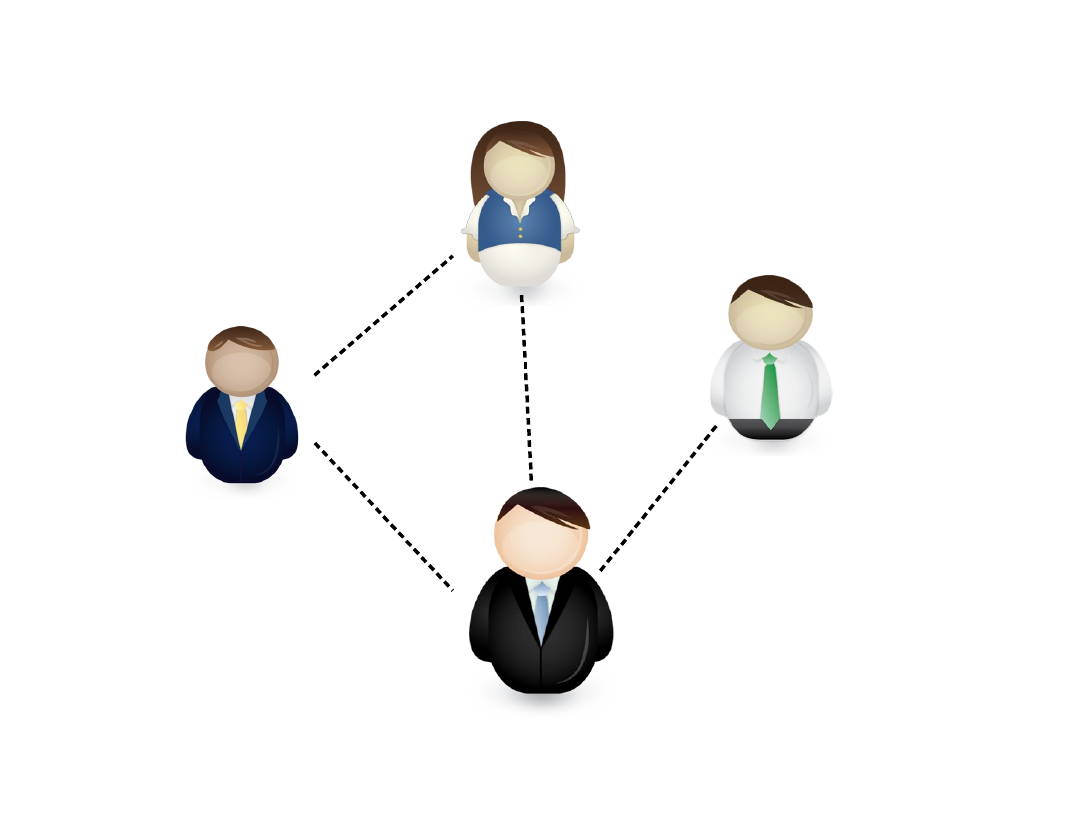}}\subfigure[User Trajectory]{\includegraphics[width=0.45\columnwidth]{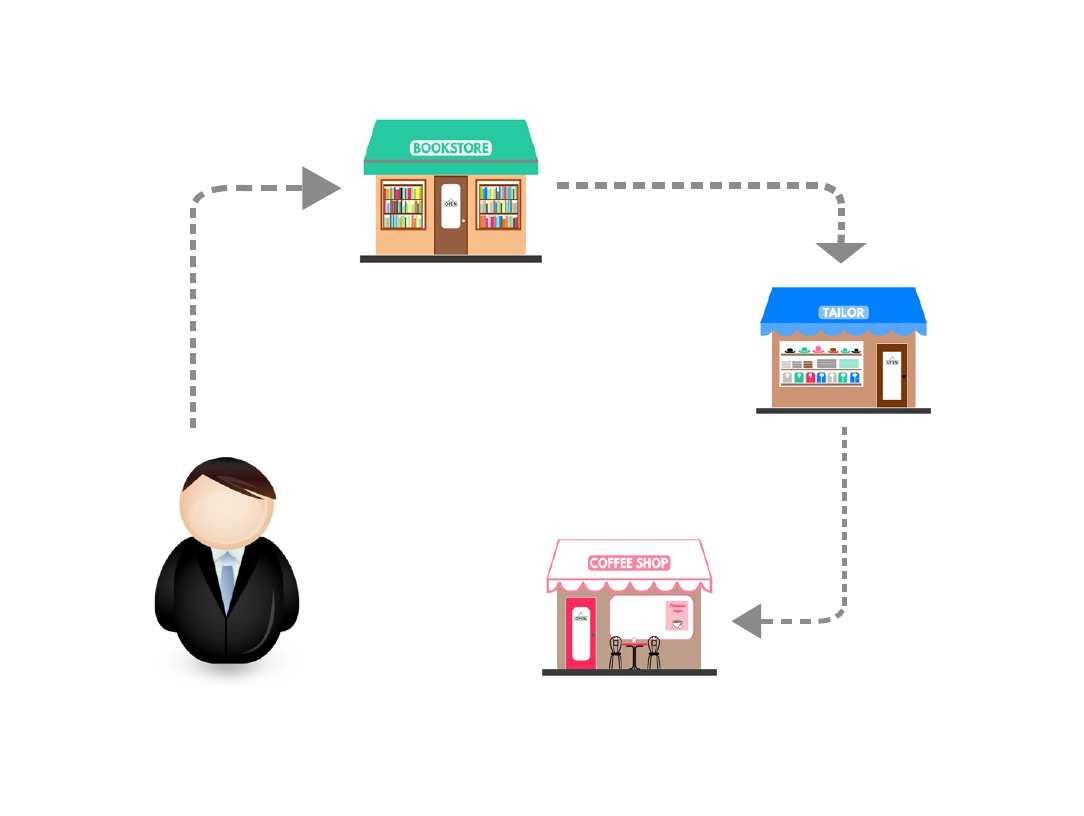}}
\end{minipage}
\caption{An illustrative example for the data in LBSNs: (a) Link connections represent the friendship between users. (b) A trajectory generated by a user is a sequence of chronologically ordered check-in records.}
\label{fig:exp}
\end{figure}

Apart from recording user trajectory data, another major feature of these location-based services is that they also allow the users to connect whomever they like or are interested in. For example, with \emph{Brightkite} you can track on your friends or any other Brightkite users nearby using the phone's built in GPS. A combination of social networking and location-based services has lead to a specific style of social networks, termed as  \emph{location-based social networks (LBSN)}~\cite{cho2011friendship,bao2012location,zheng2015trajectory}. We present an illustrative example for LBSNs in Fig.~\ref{fig:exp}, and it can been seen that LBSNs usually include both the social network and mobile trajectory data.
Recent literature has shown that social link information is useful to improve existing recommendation tasks~\cite{machanavajjhala2011personalized,yuan2014generative,ma2014measuring}.  Intuitively, users that often visit the same or similar locations are likely to be social friends\footnote{Note that online social relationship does not necessarily indicate offline friendship in real life. } and social friends are likely to visit same or similar locations.
Specially, several studies have found that there exists close association between social connections and trajectory behaviors of users in LBSNs.
 On one hand, as shown in \cite{cho2013socially}, locations that are frequently visited by socially-related persons tend to be correlated.
 On the other hand,  trajectory similarity can be utilized to infer social strength between users~\cite{pham2013ebm,zheng2010geolife,zheng2011recommending}.  Therefore we need to develop a comprehensive view to analyze and mine the information from the two aspects. In this paper, our focus is to develop a joint approach to model LBSN data by characterizing both the social network and mobile trajectory data.

\emph{In the first aspect,} social network analysis has attracted increasing attention during the past decade.
It characterizes network structures in terms of nodes (individual actors, people, or things within the network) and the ties or edges (relationships or interactions) that connect them.
A variety of applications have been developed on social networks, including network classification~\cite{sen2008collective}, link prediction~\cite{liben2007link}, anomaly detection~\cite{chandola2009anomaly} and community detection~\cite{fortunato2010community}. A fundamental  issue is how to represent network nodes. Recently, networking embedding models \cite{Perozzi:2014:DOL:2623330.2623732} have been proposed in order to solve the data sparsity in networks.
\emph{In the second aspect},  location-based services  provide a convenient way for users to record their trajectory information, usually called \emph{check-in}. Independent of social networking analysis, many studies have been constructed to improve the location-based services.
A typical application task is the location recommendation, which aims to infer users' visit preference and make meaningful recommendations for users to visit. It can be divided into three different settings:  general location recommendation  \cite{zheng2009mining,cheng2012fused,ye2011exploiting}, time-aware location recommendation \cite{yuan2013time,yuan2014graph,liu2016predicting} and next-location recommendation\cite{cheng2013you,ye2013s,zhang2014lore}.
General location recommendation will generate  an overall recommendation list of locations for a users to visit; while time-aware or next location recommendation further imposes the temporal constraint on the recommendation task by either specifying the time period or producing sequential predictions.

These two aspects capture different data characteristics on LBSNs and tend to be correlated with each other \cite{cheng2012fused,levandoski2012lars}.
To conduct better and more effective data analysis and mining studies, there is a need to develop a joint model by capturing both network structure and trajectory behaviors on LBSNs. However, such a task is challenging. Social networks and mobile trajectories are heterogeneous data types. A social network is typically characterized  by a graph, while a trajectory is usually modelled as a sequence of check-in records. A commonly used way to incorporate social connections into an application system (\eg recommender systems) is to adopt the regularization techniques  by assuming that the links convey the user similarity. In this way, the social connections are exploited as the side information but not characterized by a joint data model,  and the model performance highly rely on the ``homophily principle" of \emph{like associates with like}. In this paper, we take the initiative to jointly model social networks and mobile trajectories using a neural network approach. Our approach is inspired by the recent progress on deep learning.  Compared with other methods, neural network models can serve as an  effective and general function approximation mechanism that is able to capture complicated data characteristics~\cite{mittal2016survey}. In specific, recent studies have shown the superiority of neural network models on network and sequential data. First,
several pioneering studies try to embed vertices of a network into low-dimensional vector spaces~\cite{tang2011leveraging,Perozzi:2014:DOL:2623330.2623732,tang2015line}, called \emph{networking embedding}. With such a low-dimensional dense vector,  it can alleviate the data sparsity that  a sparse network representation suffers from. Second,  neural network models are powerful computational data models that are able to capture and represent complex input/output relationships. Especially, several neural network models
for processing sequential data have been proposed, such as recurrent neural networks (RNN) \cite{mikolov2010recurrent}. RNN and its variants including LSTM and GRU have shown good performance in many applications.

By combining the merits from both network embedding and sequential modelling from deep learning, we present a novel neural network model which can jointly model both social networks and mobile trajectories. In specific, our model consists of two components: the construction of social networks and the generation of mobile trajectories. We first adopt a network embedding method for the construction of social networks: a networking representation can be derived for a user. The key of our model lies in the component generating mobile trajectories. We have considered four factors that influence the generation process of mobile trajectories, namely user visit preference, influence of friends, short-term sequential contexts and long-term sequential contexts. The first two factors are mainly related to the users themselves, while the last factors mainly reflect the sequential characteristics of historical trajectories. We set two different user representations to model the first two factors: a visit interest representation and a network representation. To characterize the last two contexts, we employ the RNN and GRU models to capture the sequential relatedness in mobile trajectories at different levels, \ie short term or long term. Finally, the two components are tied by sharing the user network representations: the information from the network structure
is encoded in the user networking representation, which is subsequently utilized in the generation process of mobile trajectories.

To demonstrate the effectiveness of the proposed model, we evaluate our model using real-world datasets on two important LBSN applications, namely \emph{next-location recommendation} and \emph{friend recommendation}. For the first task, the trajectory data is the major information signal while network structure serves as auxiliary data. Our method consistently outperforms several competitive baselines. Interestingly, we have found that for users with little check-in data, the auxiliary data (\ie network structure) becomes more important to consider. For the second task, the network data is the major information signal while trajectory data serves as auxiliary data. The finding is similar to that in the first task: our method still performs best, especially for those users with few friend links.
Experimental results on the two important applications demonstrate the effectiveness of our model. In our approach, network structure and trajectory information complement each other. Hence, the improvement over baselines is more significant when either network structure or trajectory data is sparse.

Our contributions are three-fold summarized below:

\begin{itemize}
\item We proposed a novel neural network model to jointly characterize social network structure and users' trajectory behaviors. In our approach, network structure and trajectory information complement each other. It provides a promising way to characterize heterogeneous data types in LBSNs.
\item Our model considered four factors in the generation of mobile trajectories, including user visit preference, influence of friends, short-term sequential contexts and long-term sequential contexts. The first two factors are modelled by two different embedding representations for users. The model further employed both RNN and GRU models to capture both short-term and long-term sequential contexts.
\item Experimental results on two important applications demonstrated the effectiveness of our model. Interestingly, the improvement over baselines was more significant when either network structure or trajectory information was sparse.
\end{itemize}

The remainder of this paper is organized as follows. Section 2 reviews the related work and Section 3 presents the problem formulation.
The proposed model together with the learning algorithm is given in Section 4. We present the experimental evaluation in Section 5.
Section 6 concludes the paper and presents the future work.

\section{Related Work}
Our work is mainly related to distributed representation learning, social link prediction and location recommendation.
\subsection{Distributed Representation Learning and Neural Network Models}
Machine learning algorithms based on data representation learning make a great success in the past few years. Representations learning of the data can extract useful information for learning classifiers and other predictors. Distributed representation learning has been widely used in many machine learning tasks \cite{bengio2013representation}, such as computer vision \cite{krizhevsky2012imagenet} and natural language processing \cite{mikolov2013distributed}.

 During the last decade, many works have also been proposed for network embedding learning \cite{chen2007directed,tang2009relational,tang2011leveraging,Perozzi:2014:DOL:2623330.2623732}. Traditional network embedding learning algorithms learn vertex representations by computing eigenvectors of affinity matrices~\cite{belkin2001laplacian,yan2007graph,tang2011leveraging}. For example, DGE \cite{chen2007directed} solves generalized eigenvector computation problem on combinational Laplacian matrix; SocioDim~\cite{tang2011leveraging} computes $k$ smallest eigenvectors of normalized graph Laplacian matrix as $k$-dimensional vertex representations.

DeepWalk~\cite{Perozzi:2014:DOL:2623330.2623732} adapts Skip-Gram~\cite{mikolov2013distributed}, a widely used language model in natural language processing area, for NRL on truncated random walks. DeepWalk which leverages deep learning technique for network analysis is much more efficient than traditional NRL algorithms and makes large-scale NRL possible. Following this line, LINE~\cite{tang2015line} is a scalable network embedding algorithm which models the first-order and second-order proximities between vertices and GraRep~\cite{cao2015grarep} characterizes local and global structural information for network embedding by computing SVD decomposition on $k$-step transition probability matrix. MMDW~\cite{tumax} takes label information into account and learn semi-supervised network embeddings.

TADW~\cite{yang2015network} and PTE~\cite{tang2015pte} extend DeepWalk and LINE by incorporating text information into NRL respectively. TADW embeds text information into vertex representation by matrix factorization framework and PTE learns semi-supervised embeddings from heterogeneous text networks. However both TADW and PTE conduct experiments on document networks and fail to take sequential information between words into consideration.

Neural network models have achieved great success during the last decade. Two well-know neural network architectures are Convolutional Neural Network (CNN) and Recurrent Neural Network (RNN). CNN is used for extracting fix length representation from various size of data \cite{krizhevsky2012imagenet}. RNN and its variant GRU which aim at sequential modeling have been successfully applied in sentence modeling \cite{mikolov2010recurrent}, speech signal modeling \cite{chung2014empirical} and sequential click prediction \cite{zhang2014sequential}.
\subsection{Social Link Prediction}
Social link prediction has been widely studied in various social networks by mining graph structure patterns such as triadic closure process \cite{romero2010directed} and user demographics \cite{huang2014mining}. In this paper, we mainly focus on the applications on trajectory data.

Researchers used to measure user similarity by evaluating sequential patterns. For example, they used a sequence of stay points to represent a user trajectory and evaluated user similarity by a sequence matching algorithm \cite{li2008mining}. In order to improve these methods, people also took pre-defined tags and weights into consideration to better characterize stay points \cite{xiao2010finding}. As LBSN becomes increasingly popular, trajectory similarity mining has attracted much more attention. A number of factors was considered to better characterize the similarity. As a result, physical distance \cite{cranshaw2010bridging}, location category \cite{lee2011user}, spatial or temporal co-location rate \cite{wang2011human} and co-occurrence with time and distance constraints \cite{pham2011towards} were proposed for social link prediction. The diversity of co-occurrence and popularity of locations \cite{pham2013ebm} were proved to be important features among all the factors. Using associated social ties to cluster locations, the social strength can be inferred in turn by extracted clusters shared by users \cite{cho2013socially,zheng2011recommending}.
\subsection{Location Recommendation}

One of the most important tasks on trajectory modeling is location recommendation. For general location recommendation, several kinds of side information are considered, such as geographical \cite{cheng2012fused,ye2011exploiting}, temporal~\cite{zhao2016probabilistic} and social network information \cite{levandoski2012lars}. To address the data sparsity issue, content information including location category labels is also concerned \cite{yin2013lcars,zhougeneral}. The location labels and tags can also be used in probabilistic model such as aggregate LDA \cite{gao2015content}. Textual information which includes text descriptions \cite{gao2015content,li2010contextual,zhao2015sar} are applied for location recommendation as well. $W^4$ employs tensor factorization on multi-dimensional collaborative recommendation for Who (user), What (location category), When(time) and Where (location) \cite{zheng2010collaborative,bhargava2015and}. However, these methods which are mainly based on collaborate filtering, matrix factorization or LDA do not model the sequential information in the trajectory.

For time-aware location recommendation task which recommends locations at a specific time, it is also worth modeling the temporal effect. Collaborate filtering based method \cite{yuan2013time} unifies temporal and geographical information with linear combination. Geographical-temporal graph was proposed for time-aware location recommendation by doing preference propagation on the graph \cite{yuan2014graph}. In addition, temporal effect is also studied via nonnegative matrix factorization \cite{gao2013exploring} and RNN \cite{liu2016predicting}.

Different from general location recommendation, next-location recommendation also need to take current state into account. Therefore, the sequential information is more important to consider in next location recommendation. Most previous works model sequential behaviors, \ie trajectories of check-in locations, based on Markov chain assumption which assumes the next location is determined only by current location and independent of previous ones~\cite{rendle2010factorizing,cheng2013you,ye2013s,zhang2014lore}. For example, Factorized Personalized Markov Chain (FPMC) algorithm~\cite{rendle2010factorizing} factorizes the tensor of transition cube which includes transition probability matrices of all users. Personalized Ranking Metric Embedding (PRME)~\cite{feng2015personalized} further extends FPMC by modeling user-location distance and location-location distance in two different vector spaces. Hierarchical Representation Model (HRM)~\cite{wang2015learning}, which is originally designed for user purchase behavior modeling, can be easily adapted for modeling user trajectories. HRM builds a two-layer structure to predict items in next transaction with user features and items in last transaction. These methods are applied for next-location recommendation which aims at predicting the next location that a user will visit, given check-in history and current location of the user. Note that Markov chain property is a strong assumption that assumes next location is determined only by current location. In practice, next location may also be influenced by the entire check-in history.

\section{Problem Formalization}
We use $L$ to denote the set of locations (a.k.a. \emph{check-in} points or POIs) .
When a user $v$ checks in at a location $l$ at the timestamp $s$, the information can be modeled as a triplet $\langle v, l, s \rangle$.  Given a user $v$, her trajectory $T_v$ is a sequence of  triplets related to $v$:  $\langle v, l_1, s_1 \rangle, ..., \langle v, l_i, s_i \rangle, ..., \langle v, l_{N}, s_{N} \rangle$, where $N$ is the sequence length and the triplets are ordered by timestamps ascendingly. For brevity, we rewrite the above formulation of $T_v$ as a sequence of locations $T_v=\{l_1^{(v)},l_2^{(v)},\dots,l_N^{(v)}\}$ in chronological order. Furthermore, we can split a trajectory into multiple consecutive subtrajectories:  the trajectory $T_v$ is split into $m_v$ subtrajectories $T_v^1,\dots,T_v^{m_v}$. Each subtrajectory is essentially a subsequence of the original trajectory sequence.  In order to split the trajectory, we compute the time interval between two check-in points in the original trajectory sequence,
we follow \cite{cheng2013you} to make a splitting when the time interval is larger than six hours.  To this end, each user corresponds to a trajectory sequence $T_v$
consisting of several consecutive subtrajectories $T_v^1,\dots,T_v^{m_v}$. Let $T$ denote the set of trajectories for all the users.

Besides trajectory data, location-based services provide social connection links among users, too.
Formally, we model the social network as a graph $G=(V,E)$, where each vertex $v\in V$ represents a user, each edge $e\in E$ represents the friendship between two users. In real applications, the edges can be either undirected or directed. As we will see, our model is flexible to deal with both types of social networks. Note that these links mainly reflect online friendship, which do not necessarily indicate that two users are friends in actual life.

Given the social network information $G=(V,E)$ and the mobile trajectory information $T$, we aim to develop a joint model which can characterize and utilize both kinds of data resources. Such a joint model should be more effective those built with a single data resource alone. In order to test the model performance, we set up two application tasks in LBSNs.

\paratitle{Task I.} For the task of next-location recommendation, our goal is to recommend  a ranked list of locations that a user $v$ is likely to visit next at each step.

\paratitle{Task II.} For the task of friend recommendation, our goal is to recommend a ranked list of users that are likely to be the friends of a user $v$.

We select these tasks because they are widely studied in LBSNs, respectively representing two aspects for mobile trajectory mining and social networking analysis.  Other tasks related to LBSN data can be equally solved by our model, which are not our focus in this paper.


\section{The Proposed Model}

In this section, we present a novel neural network model for generating both social network and mobile trajectory data.
In what follows, we first study how to characterize each individual component.  Then, we present the joint model followed by the parameter learning algorithm.
Before introducing the model details, we first summarize the used notations in this paper in Table~\ref{tab:notations}.


\begin{table}
\tbl{Notations used in this paper.\label{tab:notations}}{
\begin{tabular}{c|c}
\hline
 Notation & Descriptions\\
\hline
\hline
$V, E$ & vertex and edge set \\
\hline
$L$ & location set \\
\hline
$T_v, T_v^j$ & trajectory and the $j$-th subtrajectory of user $v$\\
\hline
$m_v$ & number of subtrajectories in the trajectory $T_v$ of user $v$\\
\hline
$m_{v,j}$ & number of locations in the $j$-th subtrajectory of trajectory $T_v$ of user $v$\\
\hline
$l_i^{(v,j)}$ & the $i$-th location of the $j$-th subtrajectory of user $v$\\
\hline
$U_{l_i}$ & representation of location $l_i$ used in representation modeling\\
\hline
$U'_{l_i}$ & representation of location $l_i$ for prediction\\
\hline
$P_v, F_{v}$ &interest and friendship representation of user $v$\\
\hline
$F'_{v}$ &context friendship representation of user $v$\\
\hline
$S_i$ & short-term context representation after visiting location $l_{i-1}$\\
\hline
$h_t$ & long-term context representation after visiting location $l_{t-1}$\\
\hline
\end{tabular}}
\end{table}


\subsection{Modeling the Construction of the Social Network}

Recently,  networking representation learning is widely studied \cite{chen2007directed,tang2009relational,tang2011leveraging,Perozzi:2014:DOL:2623330.2623732}, and it provides a way to explore the networking structure patterns using low-dimensional embedding vectors. Not limited to discover structure patterns,  network representations have been shown to be effective to serve as important features in many network-independent tasks, such as demographic prediction \cite{huang2014mining} and text classification \cite{yang2015network}.  In our task,  we characterize the networking representations based on two considerations. First, a user is likely to have similar visit behaviors with their friends, and user links can be leveraged to share common visit patterns.  Second, the networking structure is utilized as auxiliary information to enhance the trajectory modelling.

Formally, we use a $d$-dimensional embedding vector of use $F_{v}\in \mathbb{R}^d$ to denote the network representation of user $v$ and matrix $F\in \mathbb{R}^{|V|\times d}$ to denote the network representations for all the users. The network representation is learned with the user links on the social network, and encodes the information for the structure patterns of a user.

The social network is constructed based on users' networking representations $F$. We first study how to model the generative probability for a edge of $v_i \rightarrow v_j$, formally as $\Pr[(v_i,v_j)\in E]$. The main intuition is that if two users $v_i$ and $v_j$ form a friendship link on the network, their networking representations should be similar. In other words, the inner product $F_{v_i}^\top\cdot F_{v_j}$ between the corresponding two networking representations will yield a large similarity value for two linked users.  A potential problem will be such a formulation can only deal with undirected networks. In order to characterize both undirected and directed networks,  we propose to incorporate a context representation for a user $v_j$, i.e., $F'_{v_j}$.  Given a directed link $v_i \rightarrow v_j$, we model the representation similarity as   $F_{v_i}^\top\cdot F'_{v_j}$ instead of $F_{v_i}^\top\cdot F_{v_j}$. The context representations are only used in the network construction. We define the probability of a link $v_i \rightarrow v_j$ by using a sigmoid function as follows

\begin{equation}
\Pr[(v_i,v_j)\in E] = \sigma(-F_{v_i}^\top\cdot F'_{v_j})=\frac{1}{1+\exp(-F_{v_i}^\top\cdot F'_{v_j})}.
\label{edge1}
\end{equation}
When dealing with undirected networks, a friend pair $(v_i, v_j)$  will be split into two directed links namely $v_i \rightarrow v_j$ and $v_j \rightarrow v_i$.
For edges not existing in $E$, we propose to use the following formulation

\begin{equation}
\Pr[(v_i,v_j)\not\in E] = 1-\sigma(-F_{v_i}^\top\cdot F'_{v_j})=\frac{\exp(-F_{v_i}^\top\cdot F'_{v_j})}{1+\exp(-F_{v_i}^\top\cdot F'_{v_j})}.
\label{edge2}
\end{equation}

Combining Eq.~\ref{edge1} and \ref{edge2}, we essentially adopt a Bernouli distribution for modelling networking links.
Following studies on networking representation learning \cite{Perozzi:2014:DOL:2623330.2623732}, we assume that each user pair is independent in the generation process. That is to say the probabilities $\Pr[(v_i,v_j)\in E |F]$ are independent for different pairs of $(v_i,v_j)$.  With this assumption, we can factorize the generative probabilities by user pairs

\begin{equation}
\begin{aligned}
\mathcal{L}(G)&=\sum_{(v_i,v_j)\in E}\log\Pr[(v_i,v_j)\in E]+\sum_{(v_i,v_j)\not\in E}\log\Pr[(v_i,v_j)\not\in E]\\
&=-\sum_{v_i,v_j}\log(1+\exp(-F_{v_i}^\top\cdot F'_{v_j}))-\sum_{(v_i,v_j)\not\in E}F_{v_i}^\top\cdot F'_{v_j}.
\end{aligned}
\label{network}
\end{equation}

\subsection{Modeling the Generation of the Mobile Trajectories}
In Section 3, a user trajectory is formatted as an ordered check-in sequences. Therefore, we model the trajectory generation process with a sequential neural network method. To generate a trajectory sequence, we generate the locations in it one by one conditioned on four important factors.  We first summarize the four factors as below

\begin{itemize}
\item \emph{General visit preference}: A user's preference or habits directly determine her own visit behaviors.
\item \emph{Influence of Friends}: The visit behavior of a user is likely to be influenced by her friends. Previous studies \cite{cheng2012fused,levandoski2012lars} indeed showed that socially correlated users tend to visit common locations.
\item \emph{Short-term sequential contexts}:  The next location is closely related to the last few locations visited by a user.
The idea is intuitive in that the visit behaviors of a user is usually related to a single activity or a series of related activities in a short time window, making that the visited locations have strong correlations.
\item \emph{Long-term sequential contexts}: It is likely that there exists long-term dependency for the visited locations by a user in a long time period. A specific case for long-term dependency will be periodical visit behaviors. For example, a user regularly has a travel for vocation in every summer vocation.
\end{itemize}

The first two factors are mainly related to the two-way interactions between users and locations. While the last two factors mainly reflect the sequential relatedness among the visited locations  by a user.

\subsubsection{Characterization of General Visit Preference}

We first characterize the general visit preference by the interest representations.
We use a $d$-dimensional embedding vector of $P_{v}\in \mathbb{R}^d$ to denote the visit interest representation of user $v$ and matrix $P\in \mathbb{R}^{|V|\times d}$ to denote the visit preference representations for all the users. The visit interest representation encodes the information for the general preference of a user over the set of locations in terms of visit behaviors.

We assume that one's general visit interests are relatively stable and does not vary too much in a given period. Such an assumption is reasonable in that a user typically has a fixed lifestyle (\eg with a relatively fixed residence area) and her visiting behaviors are likely to show some overall patterns. The visit interest representation aims to capture and encode such visit patterns by using a $d$-dimensional embedding vector.  For convenience, we call $P_{v}$ as the \emph{interest representation} for user $v$.

\subsubsection{Characterization of Influence of Friends}

For characterizing influence of friends,
a straightforward approach is to model the correlation between interest representations from two linked users with some regularization terms. However, such a method usually has high computational complexity. In this paper, we adopt a more flexible method: we incorporate the network representation in the trajectory generation process. Because the network representations are learned through the network links, the information from their friends  are implicitly encoded and used.  We still use the formulation of networking representation $F_v$ introduced in Section 4.1.

\subsubsection{Characterization of Short-Term Sequential Contexts}
Usually, the visited locations by a user in a short time window are closely correlated. A short sequence of the visited locations tend to be related to some activity. For example, a sequence ``Home $\rightarrow$ Traffic $\rightarrow$ Office"  refers to one's transportation activity from home to office.
 In addition, the geographical or traffic limits play an important role in trajectory generation process. For example, a user is more likely to visit a nearby location.
Therefore, when a user decides what location to visit next, the last few locations visited by herself should be of importance for next-location prediction.

\begin{figure}[htb]
\centering
\includegraphics[width=0.7\columnwidth]{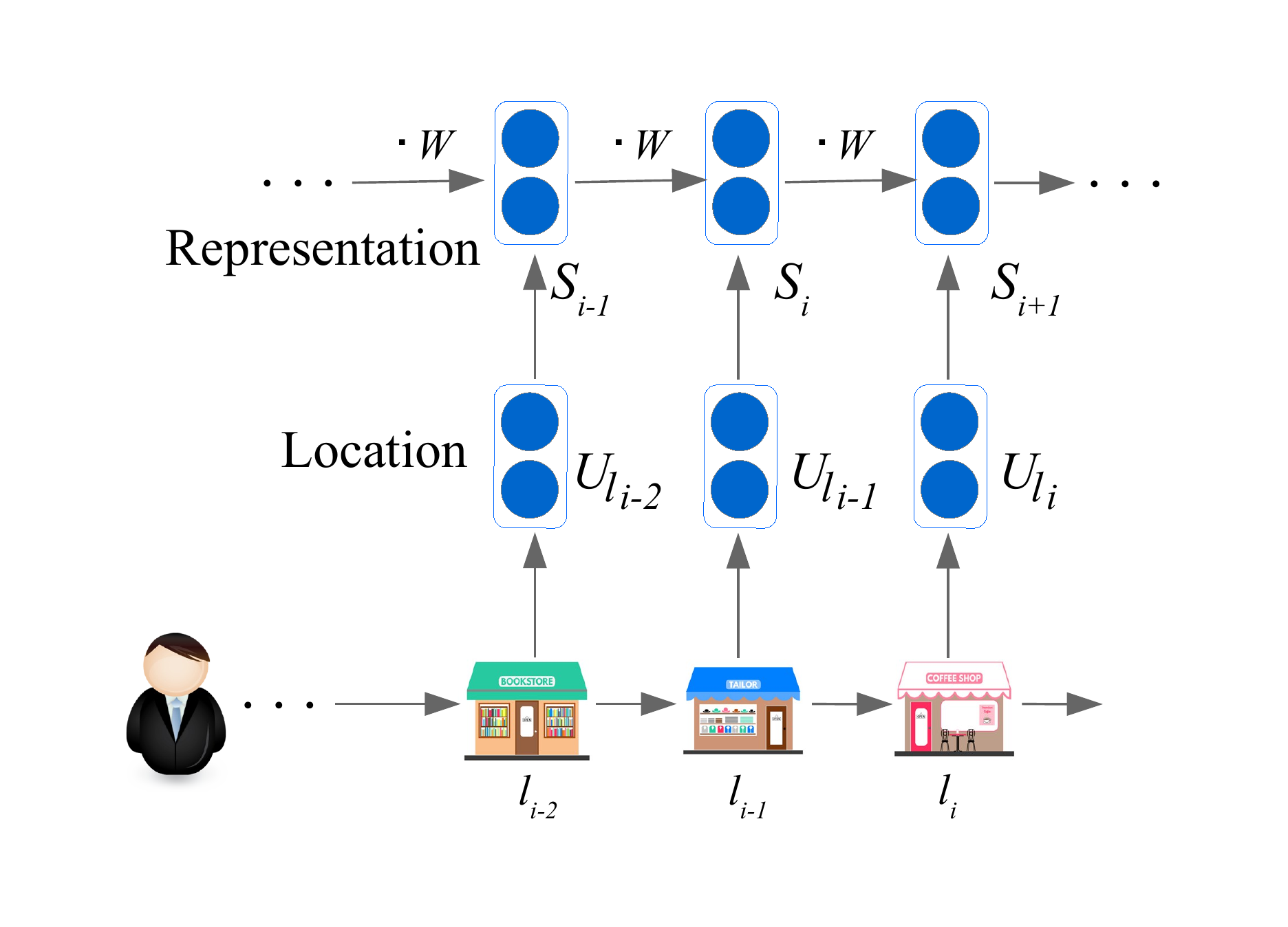}
\caption{An illustrative example of recurrent neural networks for modelling short-term sequential contexts. }
\label{fig:str}
\end{figure}

Based on the above considerations, we treat the last few visited locations in a short time window as the sequential history and predict the next location based on them.
To capture the short-term visit dependency, we use the Recurrent Neural Network (RNN), a convenient way for modelling sequential data, to develop our model.
Formally, given the $j$-th subsequence $T_v^j=\{l_1^{(v,j)},l_2^{(v,j)}\dots l_{m_{v,j}}^{(v,j)}\}$ from the trajectory of user $v$, we recursively define the short-term sequential relatedness  as follows:

\begin{equation}
S_i =\text{tanh}(U_{l_{i-1}}+W\cdot S_{i-1})
\label{state}
\end{equation}
where $S_i\in \mathbb{R}^d$ is the embedding representation for the state after visiting location $l_{i-1}$, $U_{l_i}\in \mathbb{R}^d$ is the representation of location $l_i^{(v,j)}$ and $W\in \mathbb{R}^{d\times d}$ is a transition matrix.  Here we call $S_i$ \emph{states} which are similar to those in Hidden Markov Models. RNN resembles Hidden Markov Models in that the sequential relatedness is also reflected through the transitions between two consecutive states. A major difference is that in RNN each hidden state is characterized by a $d$-dimensional embedding vector.
As shown in Fig. \ref{fig:str}, we derive the state representation $S_{i}$ by forwarding $S_{i-1}$ with a transformation matrix  $W$ and adding the embedding representation for the current location $U_{l_{i-1}}$. The initial representation $S_0$ is invariant among all users because short-term correlation is supposed to be irrelevant to user preference in our model.
Our formulation in Eq.~\ref{state} is essentially  a RNN model without outputs. The embedding vector corresponding to each state can be understood as an information summary till the corresponding location in the sequence.  Especially, the state corresponding to the last location can be considered the embedding representation for the entire sequence.



\subsubsection{Characterization of Long-Term Sequential Contexts}
In the above, short-term sequential contexts (five locations on average for our dataset) aim to capture the sequential relatedness in a short time window. The long-term sequential contexts are also important to consider when modelling trajectory sequences. For example,  a user is likely to show some periodical or long-range visit patterns.
To capture the long-term dependency,
a straightforward approach will be to use another RNN model for the entire trajectory sequence. However, the entire trajectory sequence generated by a user in a long time period tends to contain a large number of locations, \eg several hundred locations or more. A RNN model over long sequences usually suffers from the problem of ``vanishing gradient".

\begin{figure}[htb]
\centering
\includegraphics[width=0.8\columnwidth]{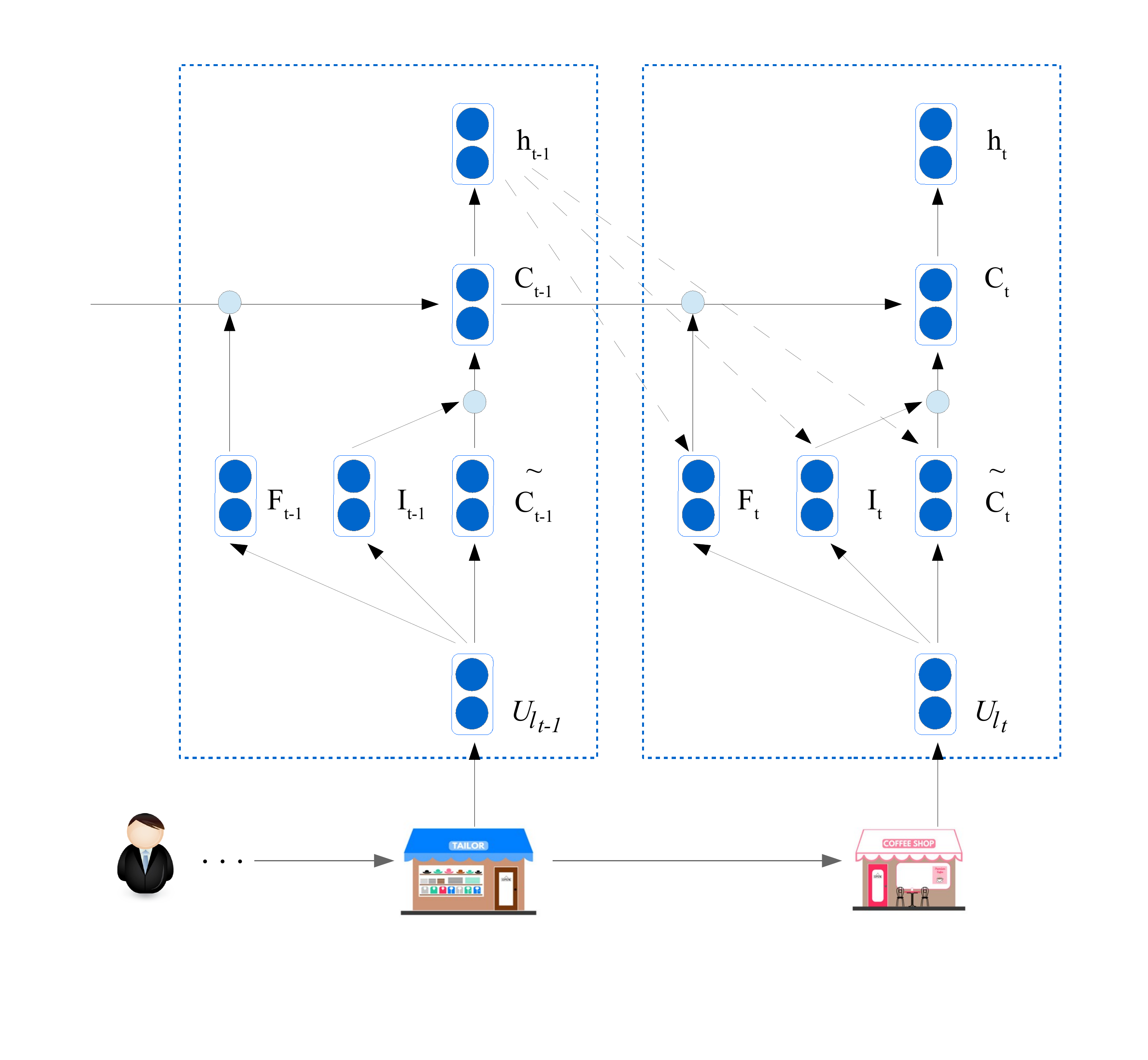}
\caption{An illustrative architecture of recurrent neural networks with GRUs.  Let $\widetilde{C_t}$ denote a candidate state. The current state $C_t$ is a mixture of the last state $C_{t-1}$ and the current candidate state $\widetilde{C_t}$. $I_t$ and $F_t$ are input and forget gate respectively, which can control this mixture.}
\label{fig:gru}
\end{figure}

To address the problem, we employ the Gated Recurrent Unit (GRU) for capturing long-term dependency in the trajectory sequence. Compared with traditional RNN, GRU incorporates several extra gates to control the input and output. Specifically, we use two gates in our model: input gate and forget gates. With the help of input and forget gates, the memory of GRU, \ie the state $C_t$ can remember the ``important stuff" even when the sequence is very long and forget less important information if necessary. We present an illustrative figure for the architecture for recurrent neural networks with GRUs in Fig.~\ref{fig:gru}.

Formally, consider the following location sequence $\{l_1,l_2,\dots,l_m\}$, we denote the initial state by $C_0\in \mathbb{R}^d$ and initial representation by $h_0=\text{tanh}(C_0)\in \mathbb{R}^d$. At a timestep of $t$, the new candidate state is updated as follows

\begin{equation}
\widetilde{C_t} =\text{tanh}(W_{c_1}U_{l_t}+W_{c_2}h_{t-1}+b_c)
\label{candgate}
\end{equation}
where $W_{c_1}\in \mathbb{R}^{d\times d}$ and $W_{c_2}\in \mathbb{R}^{d\times d}$ are the model parameters, $U_{l_t}$ is the embedding representation of location $l_t$ which is the same representation used in short-term sequential relatedness, $h_{t-1}$ is the embedding representation in the last step and $b_c\in \mathbb{R}^d$ is the bias vector. Note that the computation of $\widetilde{C_t}$ remains the same as that in RNN.

GRU does not directly replace the state with $\widetilde{C_t}$ as RNN does. Instead, GRU tries to find a balance between the last state $C_{t-1}$ and a new candidate state $\widetilde{C_t}$:
\begin{equation}
C_t=i_t*\widetilde{C_t}+f_t*C_{t-1}
\label{newstate}
\end{equation}
where $*$ is entrywise product and $i_t,f_t\in \mathbb{R}^d$ are input and forget gate respectively.

And the input and forget gates $i_t,f_t\in \mathbb{R}^d$ are defined as
\begin{equation}
i_t =\sigma(W_{i_1}U_{l_t}+W_{i_2}h_{t-1}+b_i)
\label{inputgate}
\end{equation}
and
\begin{equation}
f_t =\sigma(W_{f_1}U_{l_t}+W_{f_2}h_{t-1}+b_f)
\label{forgetgate}
\end{equation}
where $\sigma(\cdot)$ is the sigmoid function, $W_{i_1},W_{i_2}\in \mathbb{R}^{d\times d}$ and $W_{f_1},W_{f_2}\in \mathbb{R}^{d\times d}$ are input and forget gate parameters, and $b_i,b_f\in \mathbb{R}^d$ are the bias vectors.

Finally, the representation of long-term interest variation at the timestep of $t$ is derived as follows
\begin{equation}
h_t =\text{tanh}(C_t).
\label{ht}
\end{equation}

Similar to Eq.~\ref{state}, $h_t$ provides a summary which encodes the information till the $t$-th location in a trajectory sequence. We can
recursively learn the representations after each visit of a location.
\subsubsection{The Final Objective Function for Generating Trajectory Data}

Given the above discussions, we are now ready to present the objective function for generating trajectory data. Given the trajectory sequence $T_v=\{l_1^{(v)},l_2^{(v)},\dots,l_m^{(v)}\}$ of user $v$, we factorize the log likelihood according to the chain rule as follows

\begin{equation}
\begin{aligned}
\mathcal{L}(T_v)&=\log \Pr[l_1^{(v)},l_2^{(v)},\dots,l_m^{(v)} | v, \Phi]\\
&=\sum_{i=1}^m \log\Pr[l_i^{(v)}|l_1^{(v)},\dots,l_{i-1}^{(v)} ,v, \Phi],
\end{aligned}
\label{chain}
\end{equation}

where  $\Phi$ denotes all the related parameters.  As we can see, $\mathcal{L}(T_v)$ is characterized as a sum of log probabilities conditioned on the user $v$ and related parameters $\Phi$.
 Recall that the trajectory $T_v$ is split into $m_v$ subtrajectories $T_v^1,\dots,T_v^{m_v}$. Let $l_i^{(v,j)}$ denote the $i$-th location in the $j$-th subtrajectory. The contextual locations for  $l_i^{(v,j)}$ contain the preceding $(i-1)$ locations (\ie $l_1^{(v,j)}\dots l_{i-1}^{(v,j)}$) in the same subtrajectory, denoted by $l_1^{(v,j)}:l_{i-1}^{(v,j)}$, and all the locations in previous $(j-1)$ subtrajectories (\ie $T_v^1,\dots,T_v^{j-1}$ ), denoted by $T_v^1 : T_v^{j-1}$.  With these notions, we can rewrite Eq.~\ref{chain} as follows

\begin{equation}
\begin{aligned}
\mathcal{L}(T_v) &=\sum_{i=1}^m \log\Pr[l_i^{(v,j)} | \underbrace{l_1^{(v,j)}:l_{i-1}^{(v,j)}}_{\text{short-term contexts}}, \underbrace{T_v^1 : T_v^{j-1}}_{\text{long-term contexts}} , v, \Phi].
\end{aligned}
\label{chain2}
\end{equation}

Given the target location $l_i^{(v,j)}$, the term of $l_1^{(v,j)}:l_{i-1}^{(v,j)}$ corresponds to the short-term contexts, the term of $T_v^1 : T_v^{j-1}$ corresponds to the long-term contexts, and $v$ corresponds to the user context.
The key problem becomes how to model the conditional probability $\Pr[l_i^{(v,j)} | l_1^{(v,j)}:l_{i-1}^{(v,j)}, T_v^1 : T_v^{j-1} , v, \Phi]$.

\begin{figure}[htb]
\centering
\includegraphics[width=0.8\columnwidth]{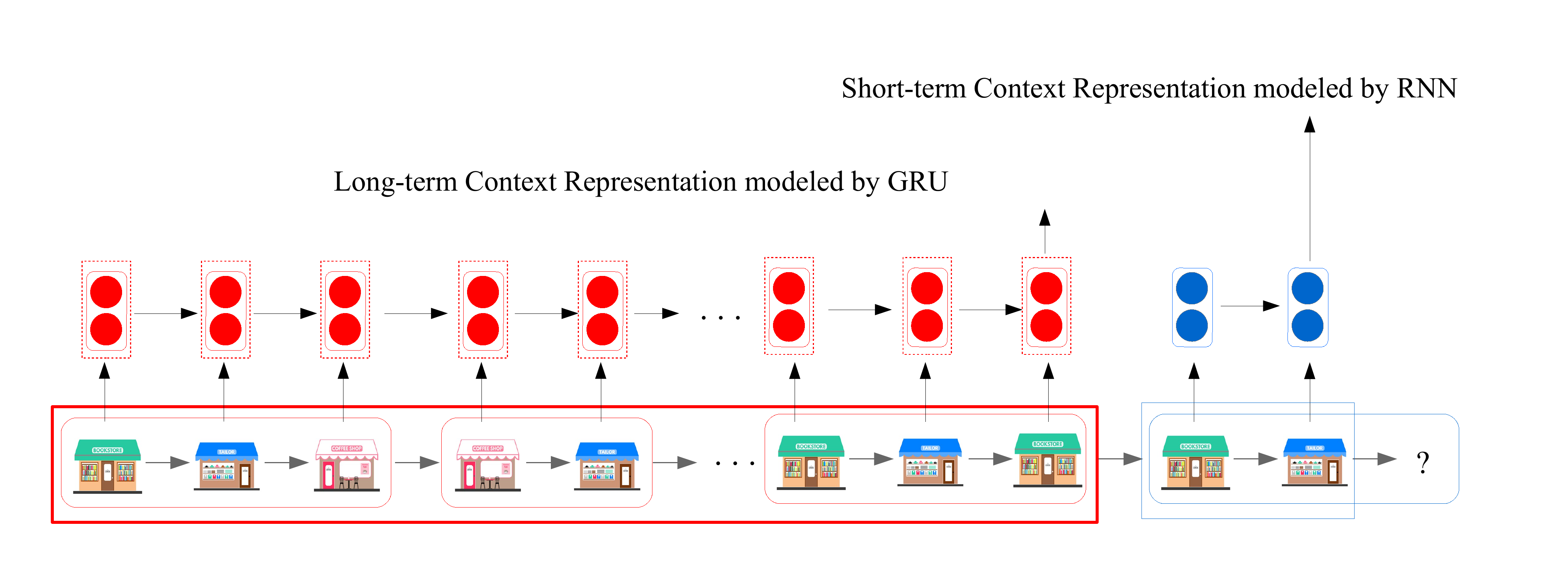}
\caption{An illustrative figure for modelling both short-term and long-term sequential contexts. The locations in a rounded rectangular indicates a subtrajectory. The locations in red and blue rectangular are used for long-term and short-term sequential contexts respectively. ``?'' is the next location for prediction.}
\label{fig:seq}
\end{figure}

For short-term contexts, we adopt the RNN model described in Eq.~\ref{state} to characterize the the location sequence of $l_1^{(v,j)}:l_{i-1}^{(v,j)}$. We use $S_{i}^j$ to denote the derived short-term representation after visiting the $i$-th location in the $j$-th subtrajectory;
For long-term contexts, the locations in the preceding subtrajectories $T_v^1\dots T_v^{j-1}$ are characterized using the GRU model in Eq.~\ref{newstate} $\sim$~\ref{ht}.
We use $h^{j}$ to denote the derived long-term representation after visiting the locations in first $j$ subtrajectories.  We present an illustrative example for the combination of short-term and long-term contexts in Fig.~\ref{fig:seq}.

So far, given a target location $l_i^{(v,j)}$, we have obtained four representations corresponding to the four factors:
networking representation (\ie $F_{v}$), visit interest representation (\ie $P_{v}$), short-term context representation $S_{i-1}^j$, and long-term context representation $h^{j-1}$. We concatenate them into a single context representation $R_v^{(i,j)}=[F_{v};P_{v};S_{i-1}^j;h^{j-1}]\in \mathbb{R}^{4d}$ and use it for next-location generation. Given the context representation $R_v^{(i,j)}$, we define the probability of $l_i^{(v,j)}$ as

\begin{eqnarray}
&&\Pr[l_i^{(v,j)}|l_1^{(v,j)}:l_{i-1}^{(v,j)}, T_v^1 : T_v^{j-1} , v, \Phi] \nonumber\\
&=&\Pr[l_i^{(v,j)}|R_v^{(i,j)}] \nonumber\\
&=& \frac{\exp(R_v^{(i,j)} \cdot U'_{l_i^{(v,j)}})}{\sum_{l\in L}\exp(R_v^{(i,j)} \cdot U'_{l})}
\label{condp}
\end{eqnarray}
where parameter $U'_{l}\in \mathbb{R}^{4d}$ is location representation of location $l\in L$ used for prediction. Note that this location representation $U'_{l}$ is totally different with the location representation $U_l\in \mathbb{R}^{d}$ used in short-term and long-term context modelling. The overall log likelihood of trajectory generation can be computed by adding up all the locations.


\subsection{The Joint Model}
Our general model is a linear combination between the objective functions for the two parts.
Given the friendship network of $G=(V,E)$ and user trajectory $T$, we have the following log likelihood function
\begin{eqnarray}
\begin{aligned}
\mathcal{L}(G,T)&= \mathcal{L}_{\text{network}}(G) + \mathcal{L}_{\text{trajectory}}(T)\\
&=\mathcal{L}(G)+\sum_{v\in V}\mathcal{L}(T_v).
\end{aligned}
\label{loglikelihood}
\end{eqnarray}

where $\mathcal{L}_{\text{network}}(G)$ is defined in Eq.~\ref{network} and  $\mathcal{L}_{\text{trajectory}}(T)=\sum_{v\in V}\mathcal{L}(T_v)$ is defined in Eq.~\ref{chain2} respectively. We name our model as \emph{Joint Network and Trajectory Model (\textbf{JNTM})}.

\begin{figure}[htb]
\centering
\includegraphics[width=0.8\columnwidth]{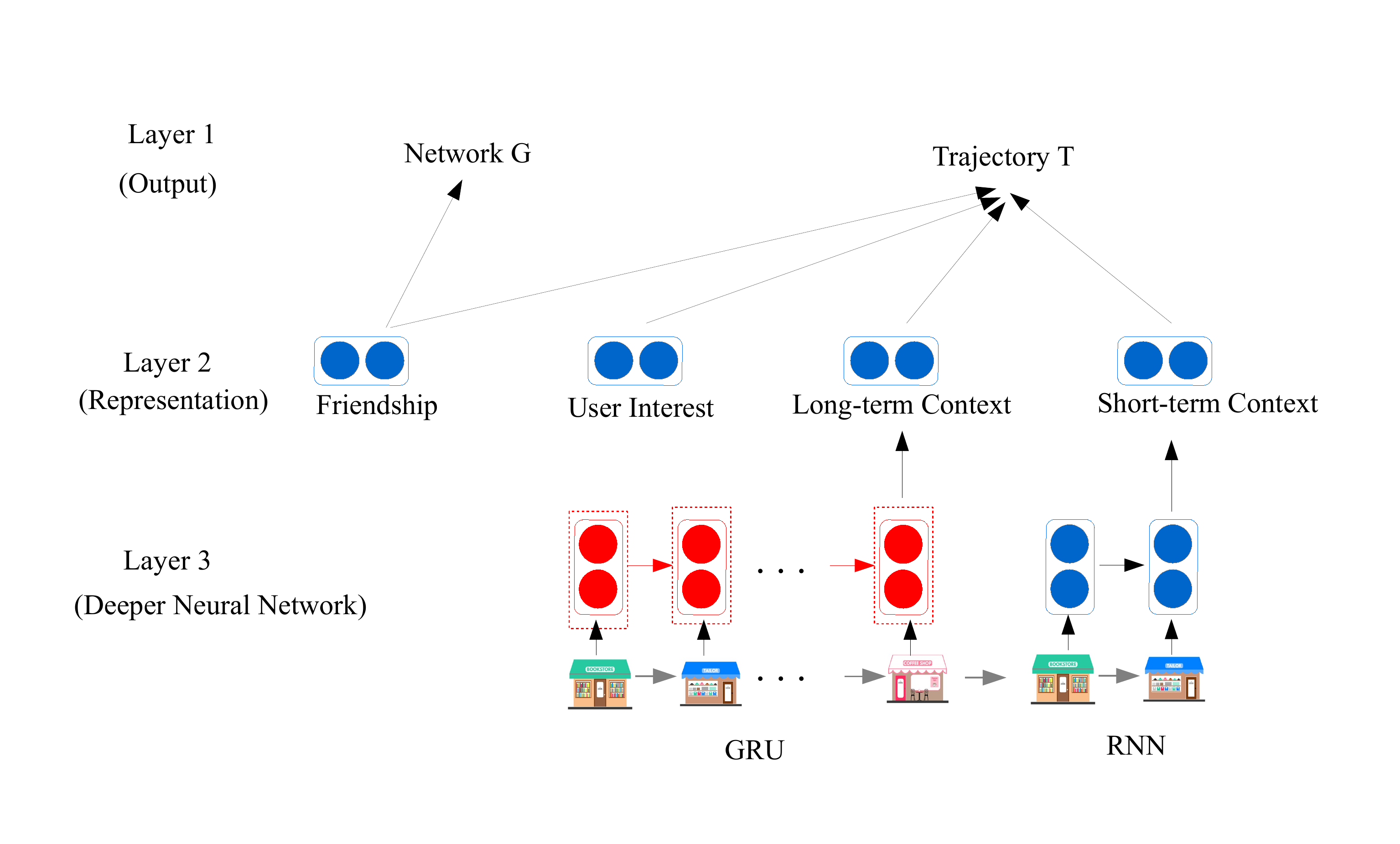}
\caption{An illustrative architecture of the proposed model JNTM.}
\label{fig:general}
\end{figure}

We present an illustrative architecture of the proposed model JNTM in Fig~\ref{fig:general}.
Our model is a three-layer neural network for generating both social network and user  trajectory.
In training, we require that both the social network and user trajectory should be provided as the objective output to the train the model.
Based on such data signals, our model naturally consists of two objective functions. For generating the  social network, a network-based user representation was incorporated; for generating the user trajectory, four factors were considered: network-based representation, general visiting preference, short-term and long-term sequential contexts.  These two parts were tied by sharing the network-based user representation.

\subsection{Parameter Learning}
Now we will show how to train our model and learn the parameters, \ie user interest representation $P\in \mathbb{R}^{|V|\times d}$, user friendship representation $F,F'\in \mathbb{R}^{|V|\times d}$, location representations $U\in \mathbb{R}^{|L|\times d},U'\in \mathbb{R}^{|L|\times 4d}$, initial short-term representation $S_{0}\in \mathbb{R}^{d}$, transition matrix $W\in \mathbb{R}^{d\times d}$, initial GRU state $C_0\in \mathbb{R}^{d}$ and GRU parameters $W_{i_1},W_{i_2},W_{f_1},W_{f_2},W_{c_1},W_{c_2}\in \mathbb{R}^{d\times d},b_i,b_f,b_c\in \mathbb{R}^{d}$ .

\paratitle{Negative Sampling.} Recall that the log likelihood of network generation equation~\ref{network} includes $|V|\times |V|$ terms. Thus it takes at least $O(|V|^2)$ time to compute, which is time-consuming. Therefore we employ negative sampling technique which is commonly used in NLP area~\cite{mikolov2013distributed} to accelerate our training process.

Note that real-world networks are usually sparse, \ie $O(E)=O(V)$. The number of connected vertex pairs (positive examples) are much less than the number of unconnected vertex pairs (negative examples). The core idea of negative sampling is that most vertex pairs serve as negative examples and thus we don't need to compute all of them. Instead we compute all connected vertex pairs and $n_1$ random unconnected vertex pairs as an approximation where $n_1\ll|V|^2$ is the number of negative samples. In our settings, we set $n_1=100|V|$. The log likelihood can be rewritten as
\begin{equation}
\mathcal{L}(G|F,F')=\sum_{(v_i,v_j)\in E}\log\Pr[(v_i,v_j)\in E]+\sum_{k=1,(v_{ik},v_{jk})\not\in E}^{n_1} \log\Pr[(v_{ik},v_{jk})\not\in E].
\label{network2}
\end{equation}
Then the computation of likelihood of network generation only includes $O(E+n_1)=O(V)$ terms.

On the other hand, the computation of equation~\ref{condp} takes at least $O(|L|)$ time because the denominator contains $|L|$ terms. Note that the computation of this conditional probability need to be done for every location. Therefore the computation of trajectory generation needs at least $O(|L|^2)$ which is not efficient. Similarly, we don't compute every term in the denominator. Instead we only compute location $l_i^{(v,j)}$ and other $n_2$ random locations. In this paper we use $n_2=100$. Then we reformulate equation~\ref{condp} as
\begin{equation}
\Pr[l_i^{(v,j)}|R_v^{(i,j)}] = \frac{\exp(R_v^{(i,j)} \cdot U'_{l_i^{(v,j)}})}{\exp(R_v^{(i,j)} \cdot U'_{l_i^{(v,j)}})+\sum_{k=1,l_k\neq l_i^{(v,j)}}^{n_2}\exp(R_v^{(i,j)} \cdot U'_{l_k})}.
\label{condp2}
\end{equation}
Then the computation of the denominator only includes $O(n_2+1)=O(1)$ terms.

 We compute the gradients of the parameters by back propagation through time (BPTT)~\cite{werbos1990backpropagation}. Then the parameters are updated with AdaGrad~\cite{duchi2011adaptive}, a variant of stochastic gradient descent (SGD), in mini-batches.

In more detail, we use pseudo codes in algorithm~\ref{alg:net} and~\ref{alg:tra} to illustrate training process of our model. The network iteration and trajectory iteration are executed iteratively until the performance on validation set becomes stable.
\begin{algorithm}[t]
\SetAlgoNoLine
\For{each user $v\in V$
    }{
    Random pick $100$ vertices $\{v_{1},\dots,v_{n_2}\}$ which are not connected with $v$\;
    Compute the log likelihood $\sum_{v':(v,v')\in E} \log \Pr[(v,v')\in E]+\sum_{k=1}^{100} \log \Pr[(v,v_k)\not\in E]$\;
    Compute the gradients of $F$ and $F'$ by back propagation\;
        }
Update $F$ and $F'$ according to the gradients\;
\caption{One Iteration of Network Generation}
\label{alg:net}
\end{algorithm}

\begin{algorithm}[t]
\SetAlgoNoLine
\For{each user $v\in V$
    }{
    Compute the forward propagation of GRU by equation~\ref{candgate}$\sim$\ref{ht} and get long-term context representations $h^1,\dots,h^{m_v}$\;
    \For{each subtrajectory $T_v^j$ of user $v$
    }{
    \For{each location $l_i^{(v,j)}$ of $T_v^j$
    }{
    Update short-term location dependency representation by equation~\ref{newstate}\;
    Concatenate four representations $[F_{v};P_{v};S_{i-1}^j;h^{j-1}]$\;
    Compute log likelihood by equation~\ref{condp2}\;
    Compute the gradients of $U',F_{v},P_{v},S_{i-1}^j$ and $h^{j-1}$
    }
    \For{$i=m_{v,j},\dots,1$
    }{
    Compute the gradients of $S_0,U,W$ through back propagation of the gradient of $S_{i-1}^j$
    }
    }
    \For{$j=m_{v},\dots,1$
    }{
    Compute the gradients of $C_0,W_{i_1},W_{i_2},W_{f_1},W_{f_2},W_{c_1},W_{c_2},b_i,b_f,b_c$ through back propagation of the gradient of $h^j$
    }
    }
    Update all parameters according to their gradients
\caption{One Iteration of Trajectory Generation}
\label{alg:tra}
\end{algorithm}

\paratitle{Complexity Analysis.} We first given the complexity analysis on time cost. The network generation of user $v$ takes $O(d)$ time to compute log likelihood and gradients of $F_v$ and corresponding rows of $F'$. Thus the complexity of network generation is $O(d|V|)$.
In trajectory generation, we denote the total number of check-in data as $|D|$. Then the forward and backward propagation of GRU take $O(d^2|D|)$ time to compute since the complexity of a single check-in is $O(d^2)$. Each step of RNN takes $O(d^2)$ time to update local dependency representation and compute the gradients of $S_0,U,W$. The computation of log likelihood and gradients of $U',F_v,P_v,S_{i-1}^j$ and $h^{j-1}$ takes $O(d^2)$ times. Hence the overall complexity of our model is $O(d^2|D|+d|V|)$. Note that the representation dimension $d$ and number of negative samples per user/location are much less than the data size $|V|$ and $|D|$. Hence the time complexity of our algorithm JNTM is linear to the data size and scalable for large datasets. Although the training time complexity of our model is relatively high, the test time complexity is small. When making location recommendations to a user in the test stage, it takes $O(d)$ time to update the hidden states of RNN/LSTM, and $O(d)$ time to evaluate a score for a single location. Usually, the hidden dimensionality $d$ is a small number, which indicates that our algorithm is efficient to make online recommendations.

In terms of space complexity, the network representations $F$ and location representations $U$ take $O((|V|+|L|)d)$ space cost in total. The space cost of other parameters is at most $O(d^2)$, which can be neglected since $d$ is much less than $|V|$ and $|L|$. Thus the space complexity of our model is similar to that of previous models such as FPMC~\cite{rendle2010factorizing}, PRME~\cite{feng2015personalized} and HRM~\cite{wang2015learning}.

\section{Experimental Evaluation}
In this section, we evaluate the performance of our proposed model JNTM. We consider two application tasks, namely next-location recommendation and friend recommendation. In what follows, we will discuss the data collection, baselines, parameter setting and evaluation metrics. Then we will present the experimental results together with the related analysis.

\subsection{Data Collection}
We consider using two publicly available LBSN datasets\footnote{http://snap.stanford.edu/data/}~\cite{cho2011friendship}, \ie Gowalla and Brightkite, for our evaluation.  Gowalla and Brightkite have released the mobile apps for users. For example, with Brightkite you can track on your friends or any other BrightKite users nearby using a phone's built in GPS; Gowalla has a similar function: use GPS data to show where you are, and what's near you.

These two datasets provide both connection links and users' check-in information. A connection link indicates reciprocal friendship and a check-in record contains the location ID and the corresponding check-in timestamp. We  organized the check-in information as trajectory sequences.  Following~\cite{cheng2013you}, we split a trajectory wherever the interval between two successive check-ins is larger than six hours. We preformed some preprocessing steps on both datasets. For \emph{Gowalla}, we removed all users who have less than $10$ check-ins and locations which have fewer than $15$ check-ins, and finally obtained $837,352$ subtrajectories. For \emph{Brightkite}, since this dataset is smaller, we only remove users who have fewer than $10$ check-ins and locations which have fewer than $5$ check-ins, and finally obtain $503,037$ subtrajectories after preprocessing.
Table \ref{tab:statistics} presents the statistics of the preprocessed datasets. Note that our datasets are larger than those in previous works~\cite{cheng2013you,feng2015personalized}.

\begin{table}[htb]
\centering
\tbl{Statistics of datasets. $|V|$: number of vertices; $|E|$: number of edges; $|D|$: number of check-ins; $|L|$: number of locations.}{
\begin{tabular}{ccccc}
\hline
 Dataset & $|V|$ & $|E|$ & $|D|$ & $|L|$\\
\hline
\hline
Gowalla & 37,800 & 390,902 & 2,212,652 & 58,410\\
Brightkite & 11,498 & 140,372 & 1,029,959 & 51,866\\
\hline
\end{tabular}}
\label{tab:statistics}
\end{table}

A major assumption we have made is that there exists close association between social links and mobile trajectory behaviors. To verify this assumption, we construct an experiment to reveal basic correlation patterns between these two factors.
For each user, we first generate a location set consisting of the locations that have been visited by the user.  Then we can measure the similarity degree between the location sets from two users using the overlap coefficient\footnote{https://en.wikipedia.org/wiki/Overlap\_coefficient}. The average overlap coefficients are $11.1\%$ and $15.7\%$ for a random friend pair (\ie two users are social friends) on Brightkite and Gowalla dataset, respectively.
As a comparison, the overlap coefficient falls to $0.5\%$ and $0.5\%$ for a random non-friend pair  (\ie two users are not social friends) on Brightkite and Gowalla dataset, respectively. This finding indicates that users that are socially connected indeed have more similar visit characteristics.
We next examine whether two users with similar trajectory behaviors are more likely to be socially connected. We have found that the probabilities that two random users are social friends are $0.1\%$ and $0.03\%$ on Brightkite and Gowalla dataset, respectively. However, if we select two users with more than $3$ common locations in their location set, the probabilities that they are social friends increase to $9\%$ and $2\%$, respectively.  The above two findings show social connections are closely correlated with mobile trajectory behaviors in LBSNs.

\subsection{Evaluation Tasks and Baselines}
\subsubsection*{Next-Location Recommendation}
For the task of next-location recommendation, we consider the following baselines:

\begin{itemize}
\item \textbf{Paragraph Vector (PV)}~\cite{Le2014Distributed} is a representation learning model for both sentence and documents using simple neural network architecture. To model trajectory data, we treat each location as a word and each user as a paragraph of location words.
\item \textbf{Feature-Based Classification (FBC)}~solves the next-location recommendation task by casting it as a multi-class classification problem. The user features are learned using DeepWalk algorithm~\cite{Perozzi:2014:DOL:2623330.2623732}, and the location features are learned using word2vec~\cite{mikolov2013distributed} algorithm (similar to the training method of PV above). These features are subsequently incorporated into a a softmax classifier, \ie a multi-class generalization of logistic regression.
\item \textbf{FPMC}~\cite{rendle2010factorizing}, which is a state-of-the-art recommendation algorithm, factorizes tensor of transition matrices of all users and predicts next location by computing the transition probability based on Markov chain assumption. It was originally proposed for product recommendation, however, it is easy adapt FPMC to deal with next-location recommendation.
\item \textbf{PRME}~\cite{feng2015personalized} extends FPMC by modeling user-location and location-location pairs in different vector spaces. PRME achieves state-of-the-art performance on next-location recommendation task.
\item \textbf{HRM}~\cite{wang2015learning} is a latest algorithm for next-basket recommendation. By taking each subtrajectory as a transaction basket, we can easily adapt HRM for next-location recommendation. It is the first study that distributed representation learning has been applied to the recommendation problem.
\end{itemize}

We select these five baselines, because they represent different recommendation algorithms. PV is based on simple neural networks, FBC is a traditional classification model using embedding features, FPMC is mainly developed in the matrix factorization framework, PRME makes specific extensions based on FPMC to  adapt to the task of next-location recommendation, and HRM adopts the distributed representation learning method for next-basket modelling.

Next, we split the data collection into the training set and test set.
The first $90\%$ of check-in subtrajectories of each user are used as the training data and the remaining  $10\%$ as test data.
To tune the parameters, we use the last $10\%$ of check-ins of training data as the validation set.

Given a user,  we predict the locations in the test set in a sequential way: for each location slot, we recommend five or ten locations to the user. For JNTM, we naturally rank the locations by the log likelihood as shown in equation~\ref{condp}. Note that negative sampling is not used in evaluation. For the baselines, we rank the locations by the transition probability for FPMC and HRM and transition distance for PRME. The predictions of PV and FBC can be obtained from the output of softmax layer of their algorithms.
 Then we report Recall@5 and Recall@10 as the evaluation metrics where Recall@K is defined as

 \begin{eqnarray}%
Recall@K=\frac{\text{\# ground truth locations in the $K$ recommended locations}}{\text{\# ground truth locations in test data}}. \nonumber
\end{eqnarray}%
Note that another common metric Precision@$K$ can be used here, too. In our experiments, we have found it is positively correlated with Recall@$K$, \ie if method $A$ has a higher Recall@$K$ score than method $B$, then method $A$ also has a higher Precision@$K$ score then method $B$. We omit the  results of Precision@$K$ for ease of presentation.

\subsubsection*{Friend Recommendation}

For the task of friend recommendation, we consider three kinds of baselines based on the used data resources, including
the method with only the networking data (\ie DeepWalk), the method with only the trajectory data (\ie PMF), and the methods with both networking and  trajectory data (\ie PTE and TADW).

\begin{itemize}
\item \textbf{DeepWalk}~\cite{Perozzi:2014:DOL:2623330.2623732} is a state-of-the-art NRL method which learns vertex embeddings from random walk sequences. It first employs the random walk algorithm to generate length-truncated random paths, and apply the word embedding technique to learn the representations for network vertices.
\item \textbf{PMF}~\cite{mnih2007probabilistic} is a general collaborative filtering method based on user-item matrix factorization. In our experiments, we build the user-location matrix using the trajectory data, and then we utilize the user latent representations for friend recommendation.
\item \textbf{PTE}~\cite{tang2015pte} develops a semi-supervised text embedding algorithm for unsupervised embedding learning by removing the supervised part and optimizing over adjacency matrix and user-location co-occurrence matrix. PTE models a conditional probability $p(v_j|v_i)$ which indicates the probability that a given neighbor of $v_i$ is $v_j$. We compute the conditional probabilities for friend recommendation.
\item \textbf{TADW}~\cite{yang2015network} further extends DeepWalk to take advantage of text information of a network. We can replace text feature matrix in TADW with user-location co-occurrence matrix by disregarding the sequential information of locations. TADW defines an affinity matrix where each entry of the matrix characterizes the strength of the relationship between corresponding users. We use the corresponding entries of affinity matrix to rank candidate users for recommendation.
\end{itemize}

To construct the evaluation collection, we randomly select $20\sim 50$ of the existing connection links as training set and leave the rest for test.
We recommend $5$ or $10$ friends for each user and report Recall@5 and Recall@10. The final results are compared by varying the training ratio from $20$ to $50$ percent. Specifically, for each user $v$,  we take all the other users who are not her friends in the training set as the candidate users.
Then,  we rank the candidate users, and recommend top $5$ or $10$ users with highest ranking scores. To obtain the ranking score of user $v_j$ when we recommend friends for user $v_i$,  DeepWalk and PMF adopt the cosine similarity between their user representations. For PTE, we use the conditional probability $p(v_j|v_i)$ which indicates the probability that a given neighbor of $v_i$ is $v_j$ as ranking scores. For TADW, we compute the affinity matrix $A$ and use the corresponding entry $A_{ij}$ as ranking scores. For our model, we rank users with highest log likelihood according to Equation \ref{edge1}.

The baselines methods and our model involves an important parameter, \ie the number of latent (or embedding) dimensions.
We use a grid search from $25$ to $100$ and set the optimal value using the validation set.  Other parameters in baselines or our model can be tuned in a similar way. For our model, the learning rate and number of negative samples are empirically set to $0.1$ and $100$, respectively. We randomly initialize parameters according to uniform distribution $U(-0.02,0.02)$.

All the experiments are executed on a $12$-core CPU server and the CPU type is Intel Xeon E5-2620 @ 2.0GHz.

\subsection{Experimental Results on Next-location Recommendation.}
Table \ref{tab:nlr} shows the results of different methods on next-location recommendation. Compared with FPMC and PRME, HRM models the
sequential relatedness between consecutive subtrajectories while the sequential relatedness in a subtrajectory is ignored. In the Brightkite dataset, the average number of locations in a subtrajectory is much less than that in the Gowalla dataset. Therefore short-term sequential contexts are more important in the Gowalla dataset and less useful in the Brightkite dataset. Experimental results in Table \ref{tab:nlr} demonstrate this intuition: HRM outperforms FPMC and PRME on Brightkite while PRME works best on Gowalla.

As shown in Table \ref{tab:nlr}, our model JNTM consistently outperforms the other baseline methods. JNTM yields $4.9\%$ and $4.4\%$ improvement on Recall@5 as compared to the best baseline HRM on the Brightkite dataset and FBC on the Gowalla dataset.
Recall that our model JNTM has considered four factors, including user preference, influence of friends, short-term  and long-term sequential contexts. All the baseline methods only characterize  user preference (or friend influence for FBC) and a single kind of sequential  contexts.  Thus, JNTM achieves the best performance on both datasets.

\begin{table}[htb]
\centering
\tbl{Results of different methods on next location recommendation.}{
\begin{tabular}{c|c|c|c|c|c|c}
\hline
Dataset   & \multicolumn{3}{c|}{Brightkite} & \multicolumn{3}{c}{Gowalla} \\ \hline
Metric (\%)  &R@1& R@5 & R@10 &R@1  & R@5 & R@10 \\ \hline
\hline
PV &18.5& 44.3 & 53.2&9.9 &27.8 &36.3 \\ 
FBC &16.7 &44.1 &54.2 &13.3& 34.4&42.3\\ 
FPMC &20.6& 45.6 & 53.8&10.1 &24.9 &31.6 \\ 
PRME &15.4 &44.6 &53.0 &12.2& 31.9&38.2\\ 
HRM  &17.4&46.2 &56.4&7.4  &26.2 &37.0\\ \hline
JNTM &$\mathbf{22.1}$ & $\mathbf{51.1}$&$\mathbf{60.3}$ & $\mathbf{15.4}$ & $\mathbf{38.8}$ &$\mathbf{48.1}$\\ \hline
\end{tabular}}
\label{tab:nlr}
\end{table}

The above results are reported by averaging over all the users. In recommender systems, an important issue is how a method performs in the cold-start setting, \ie new users or new items. To examine the effectiveness on new users generating very few check-ins, we present the results of Recall@5 for users with no more five subtrajectories in Table~\ref{tab:nllow}. In a cold-start scenario, a commonly used way to leverage the side information (\eg user links \cite{cheng2012fused} and text information \cite{gao2015content,li2010contextual,zhao2015sar}) to alleviate the data sparsity. For our model, we characterize two kinds of user representations, either using network data or trajectory data. The user representations learned using network data can be exploited to improve the recommendation performance for new users to some extent. Indeed, networking representations have been applied to multiple network-independent tasks, including profession prediction \cite{tu2015prism} or text classification \cite{yang2015network}.
By utilizing the networking representations, the results indicate that our model JNTM is very promising to deal with next-location recommendation in a cold-start setting.

\begin{table}[htb]
\centering 
\tbl{Results of next location recommendation results for users with no more than five subtrajectories.}{
\begin{tabular}{c|c|c|c|c|c|c}
\hline
Dataset   & \multicolumn{3}{c|}{Brightkite} & \multicolumn{3}{c}{Gowalla} \\ \hline
Metric (\%) &R@1& R@5 & R@10&R@1 & R@5 & R@10 \\ \hline\hline
PV &13.2& 22.0 & 26.1&4.6 &7.8 &9.2 \\ 
FBC &9.0 &29.6 &39.5 &4.9& 12.0&16.3\\ 
FPMC &17.1&30.0 &33.9  &5.5& 13.5&18.5 \\ 
PRME &22.4& 36.3& 40.0&7.2&  12.2&15.1\\ 
HRM  &12.9&31.2 &39.7 &5.2  &15.2&21.5\\ \hline
JNTM &$\mathbf{28.4}$&$\mathbf{53.7}$& $\mathbf{59.3}$ &$\mathbf{10.2}$ &$\mathbf{24.8}$ &$\mathbf{32.0}$\\ \hline
\end{tabular}}
\label{tab:nllow}
\end{table}


Note that the above experiments are based on general next-location recommendation, where we do not examine whether a recommended location has been previously visited or not by a user. To further test the effectiveness of our algorithm, we conduct experiments on next new location recommendation task proposed by previous studies~\cite{feng2015personalized}. In this setting, we only recommend new locations when the user decide to visit a place. Specifically, we rank all the locations that a user has never visited before for recommendation~\cite{feng2015personalized}. We present the experimental results in Table~\ref{tab:nlrnew}. Our method consistently outperforms all the baselines on next new location recommendation in both datasets. By combining results in Table~\ref{tab:nlr} and \ref{tab:nllow}, we can see that our model JNTM is more effective in next-location recommendation task compared to these baselines.

\begin{table}[htb]
\centering
\tbl{Results of different methods on next new location recommendation.}{
\begin{tabular}{c|c|c|c|c|c|c}
\hline
Dataset   & \multicolumn{3}{c|}{Brightkite} & \multicolumn{3}{c}{Gowalla} \\ \hline
Metric (\%)& R@1 & R@5 & R@10 & R@1 & R@5 & R@10 \\ \hline
\hline
PV &0.5& 1.5 & 2.3&1.0 &3.3 &5.3 \\ 
FBC &0.5 &1.9 &3.0 &1.0& 3.1&5.1\\ 
FPMC& 0.8 & 2.7 & 4.3 & 2.0 &6.2 & 9.9 \\ 
PRME& 0.3 & 1.1 & 1.9 & 0.6& 2.0& 3.3\\ 
HRM & 1.2& 3.5& 5.2  &1.7& 5.3& 8.2\\ \hline
JNTM &$\mathbf{1.3}$ & $\mathbf{3.7}$ & $\mathbf{5.5}$ &$\mathbf{2.7}$& $\mathbf{8.1}$ &$\mathbf{12.1}$\\ \hline
\end{tabular}}
\label{tab:nlrnew}
\end{table}

In the above, we have shown the effectiveness of the proposed model JNTM on the task of next-location recommendation. Since trajectory data itself is sequential data, our model has leveraged the flexibility of recurrent neural networks for modelling sequential data, including both short-term  and long-term sequential contexts. Now we study the effect of sequential modelling on the current task.

We prepare three variants for our model JNTM
\begin{itemize}
\item JNTM$_{base}$: it removes both short-term and long-term contexts. It only employs the user interest representation and network representation to generate the trajectory data.
\item JNTM$_{base+long}$: it incorporates the modelling for long-term contexts to JNTM$_{base}$.
\item JNTM$_{base+long+short}$: it incorporates the modelling for both short-term and long-term contexts to JNTM$_{base}$.
\end{itemize}



\begin{table}[htb]
\tbl{Performance comparison for  three variants of JNTM on next-location recommendation.}{
\begin{tabular}{c|c|c|c|c|c|c}
\hline
Dataset   & \multicolumn{3}{c|}{Brightkite} & \multicolumn{3}{c}{Gowalla} \\ \hline
Metric (\%)& R@1& R@5 & R@10& R@1  & R@5 & R@10 \\ \hline\hline
JNTM$_{base}$&20.2 & 49.3 &59.2&12.6 &36.6 & 45.5\\ \hline
JNTM$_{base+long}$&20.4 (+2\%)& 50.2 (+2\%)& 59.8 (+1\%)& 13.9 (+10\%)& 36.7 (+0\%)& 45.6 (+0\%)\\ \hline
JNTM$_{base+long+short}$ & $\mathbf{22.1} (+9\%)$& $\mathbf{51.1} (+4\%)$ &$\mathbf{60.3 (+2\%)}$ &$\mathbf{15.4 (+18\%)}$ &$\mathbf{38.8} (+6\%)$& $\mathbf{48.1} (+6\%)$\\ \hline
\end{tabular}}
\label{tab:jntm}
\end{table}

\begin{table}[htb]
\tbl{Performance comparison for  three variants of JNTM on next new location recommendation.}{
\begin{tabular}{c|c|c|c|c|c|c}
\hline
Dataset   & \multicolumn{3}{c|}{Brightkite} & \multicolumn{3}{c}{Gowalla} \\ \hline
Metric (\%)& R@1& R@5 & R@10& R@1  & R@5 & R@10 \\ \hline\hline
JNTM$_{base}$&0.8 & 2.5 &3.9&0.9 &3.3 & 5.5\\ \hline
JNTM$_{base+long}$&1.0 (+20\%)& 3.3 (+32\%)& 4.8 (+23\%)& 1.0 (+11\%)& 3.5 (+6\%)& 5.8 (+5\%)\\ \hline
JNTM$_{base+long+short}$ & $\mathbf{1.3} (+63\%)$& $\mathbf{3.7} (+48\%)$ &$\mathbf{5.5} (+41\%)$ &$\mathbf{2.7} (+200\%)$ &$\mathbf{8.1} (+145\%)$& $\mathbf{12.1} (+120\%)$\\ \hline
\end{tabular}}
\label{tab:jntmnew}
\end{table}

Table~\ref{tab:jntm} and~\ref{tab:jntmnew} show the experimental results of three JNTM variants on the Brightkite and Gowalla dataset. The numbers in the brackets indicate the relative improvement against JNTM$_{base}$. We can observe  a performance ranking: JNTM$_{base} $ $<$ JNTM$_{base+long}$ $<$ JNTM$_{base+long+short}$. The observations indicate that both kinds of sequential contexts are useful to improve the performance for next-location recommendation. In general next location recommendation (\ie both old and new locations are considered for recommendation), we can see that the improvement from short and long term context is not significant. The explanation is that a user is likely to show repeated visit behaviors (\eg visiting the locations that have been visited before), and thus user preference is more important than sequential context to improve the recommendation performance.  While for next new location recommendation, the sequential context especially short-term context yields a large improvement margin over the baseline. These results indicate that the sequential influence is more important than user preference for new location recommendation. Our finding is also consistent with previous work~\cite{feng2015personalized}, \ie sequential context is important to consider for next new location recommendation.

 \subsection{Experimental Results on Friend Recommendation}
 \begin{table*}[htb]
\centering
\tbl{Friend recommendation results on Brightkite dataset.}{
\begin{tabular}{c|r|r|r|r|r|r|r|r}
\hline
Training Ratio   & \multicolumn{2}{c|}{20\%} & \multicolumn{2}{c|}{30\%} & \multicolumn{2}{c|}{40\%} & \multicolumn{2}{c}{50\%}\\ \hline
Metric (\%) & R@5 & R@10 & R@5 & R@10 & R@5 & R@10 & R@5 & R@10\\ \hline\hline
DeepWalk &2.3 &3.8 &3.9 &6.7 &5.5 &9.2 &7.4 &12.3 \\ 
PMF & 2.1&3.6 &2.1 &3.7 &2.3 &3.4 &2.3 &3.8\\ 
PTE & 1.5&2.5 &3.8 &4.7 &4.0 &6.6 &5.1 &8.3\\ 
TADW & 2.2& 3.4& 3.6&3.9 &2.9 &4.3 &3.2 &4.5\\ \hline
JNTM &$\mathbf{3.7}$ &$\mathbf{6.0}$ &$\mathbf{5.4}$ &$\mathbf{8.7}$ &$\mathbf{6.7}$ &$\mathbf{11.1}$ &$\mathbf{8.4}$ &$\mathbf{13.9}$\\ \hline
\end{tabular}}
\label{tab:frb}
\end{table*}

\begin{table*}[htb]
\centering
\tbl{Friend recommendation results on Gowalla dataset.}{
\begin{tabular}{c|r|r|r|r|r|r|r|r}
\hline
Training Ratio   & \multicolumn{2}{c|}{20\%} & \multicolumn{2}{c|}{30\%} & \multicolumn{2}{c|}{40\%} & \multicolumn{2}{c}{50\%}\\ \hline
Metric (\%) & R@5 & R@10 & R@5 &R@10 & R@5 & R@10 & R@5 & R@10\\ \hline\hline
DeepWalk &2.6 &3.9 &5.1 &8.1 &$\mathbf{7.9}$ &$\mathbf{12.1}$ &$\mathbf{10.5}$ &$\mathbf{15.8}$ \\ 
PMF & 1.7 & 2.4 &1.8 &2.5 &1.9 &2.7 &1.9 &3.1\\ 
PTE & 1.1&1.8 & 2.3&3.6 &3.6 &5.6 &4.9 &7.6\\ 
TADW & 2.1 &3.1 &2.6 &3.9 &3.2 &4.7 &3.6 &5.4\\ \hline
JNTM & $\mathbf{3.8}$&$\mathbf{5.5}$ & $\mathbf{5.9}$&$\mathbf{8.9}$&$\mathbf{7.9}$&11.9 &10.0 &15.1 \\ \hline
\end{tabular}}
\label{tab:frg}
\end{table*}

 We continue to present and analyze the experimental results on the task of friend recommendation.
 Table \ref{tab:frg} and \ref{tab:frb} show the results  when the training ratio varies from $20\%$ to $50\%$.

Among the baselines, DeepWalk performs best and even better than the baselines using both networking data and trajectory data (\ie PTE and TADW).  A major reason is that DeepWalk is tailored to the reconstruction of network connections and adopts a distributed representation method to capture the topology structure. As indicated in other following studies \cite{Perozzi:2014:DOL:2623330.2623732,tang2015line}, distributed representation learning is particularly effective to network embedding.
Although PTE and TADW utilize both network and trajectory data, their performance is still low. These two methods cannot capture the sequential relatedness in trajectory sequences.
Another  observation is that PMF (\ie factorizing the user-location matrix) is better than PTE at the ratio of $20\%$ but becomes the worst baseline. It is because that PMF learns user representations using the trajectory data, and the labeled data (\ie links) is mainly used for training a classifier.


Our algorithm is competitive with state-of-the-art network embedding method DeepWalk and outperforms DeepWalk when network structure is sparse. The explanation is that trajectory information is more useful when network information is insufficient. As network becomes dense, the trajectory information is not as useful as the connection links.
To demonstrate this explanation, we further report the results for users with fewer than five friends when the training ratio of $50\%$. As shown in Table \ref{tab:frlow}, our methods have yielded $2.1\%$ and $1.5\%$ improvement than DeepWalk for these inactive users on the Brightkite and Gowalla datasets, respectively. The results indicate that trajectory information is useful to improve the performance of friend recommendation for users with very few friends.

\begin{table}[htb]
\tbl{Friend recommendation results for users with fewer than five friends when training ratio is 50\%.}{
\begin{tabular}{c|c|c|c|c}
\hline
Dataset   & \multicolumn{2}{c|}{Brightkite} & \multicolumn{2}{c}{Gowalla} \\ \hline
Metric (\%)& R@5 & R@10  & R@5 & R@10 \\ \hline\hline
DeepWalk & 14.0 &18.6 &19.8 &23.5 \\ \hline
JNTM & $\mathbf{16.1}$ &$\mathbf{20.4}$ &$\mathbf{21.3}$ &$\mathbf{25.5}$\\ \hline
\end{tabular}}
\label{tab:frlow}
\end{table}

In summary, our methods significantly outperforms existing state-of-the-art methods on both next-location prediction and friend recommendation. Experimental results on both tasks demonstrate the effectiveness of our proposed model.

\subsection{Parameter Tuning}
In this section, we study on how different parameters affect the performance of our model.
We mainly select two important parameters, \ie the number of iterations and and the number of embedding dimensions.

\begin{figure*}[ht]
\centering
\begin{minipage}{\textwidth}
\subfigure[Network log likelihood]{
\includegraphics[width=0.3\textwidth]{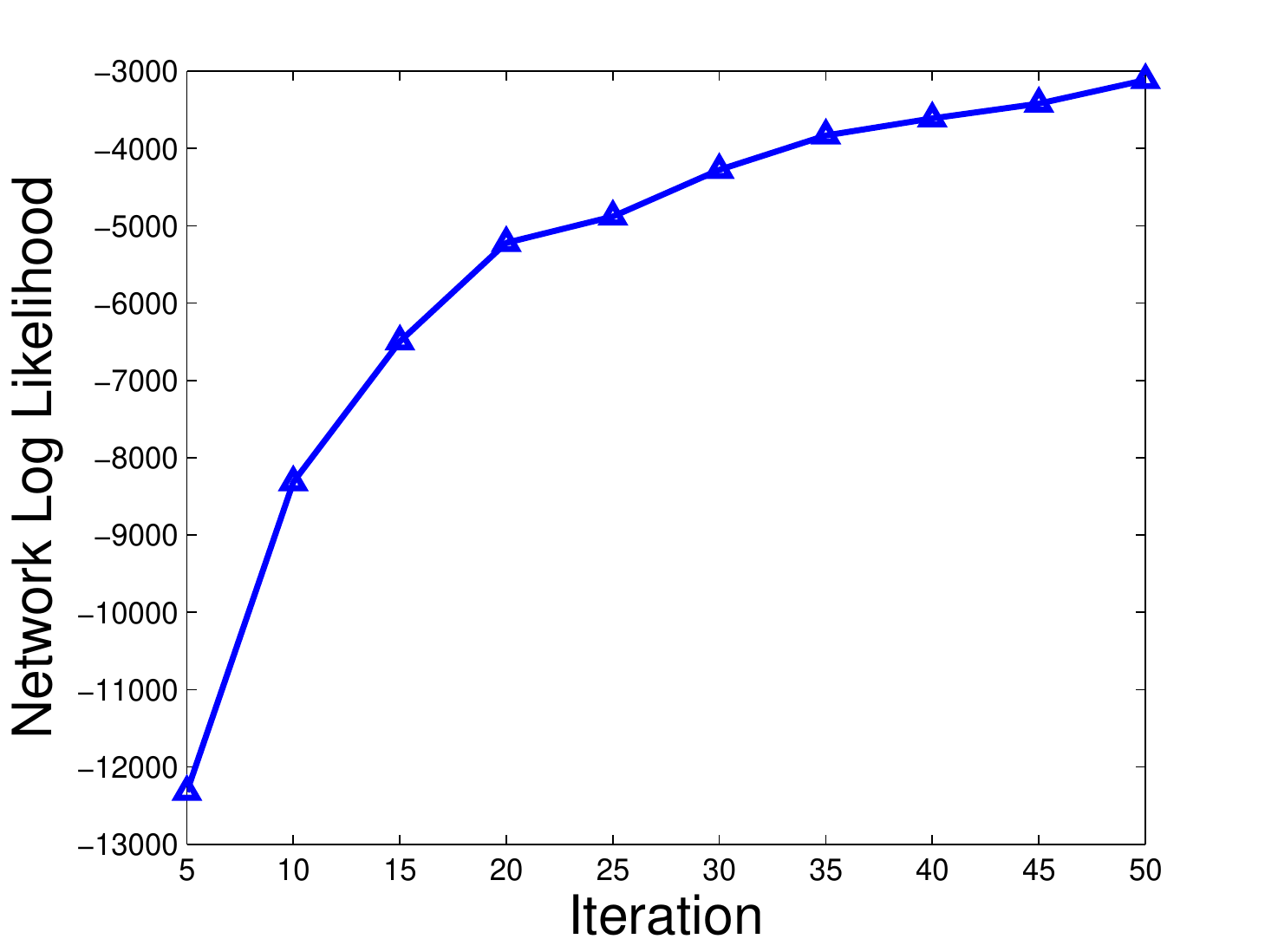}
}
\subfigure[Trajectory log likelihood.]{
\includegraphics[width=0.3\textwidth]{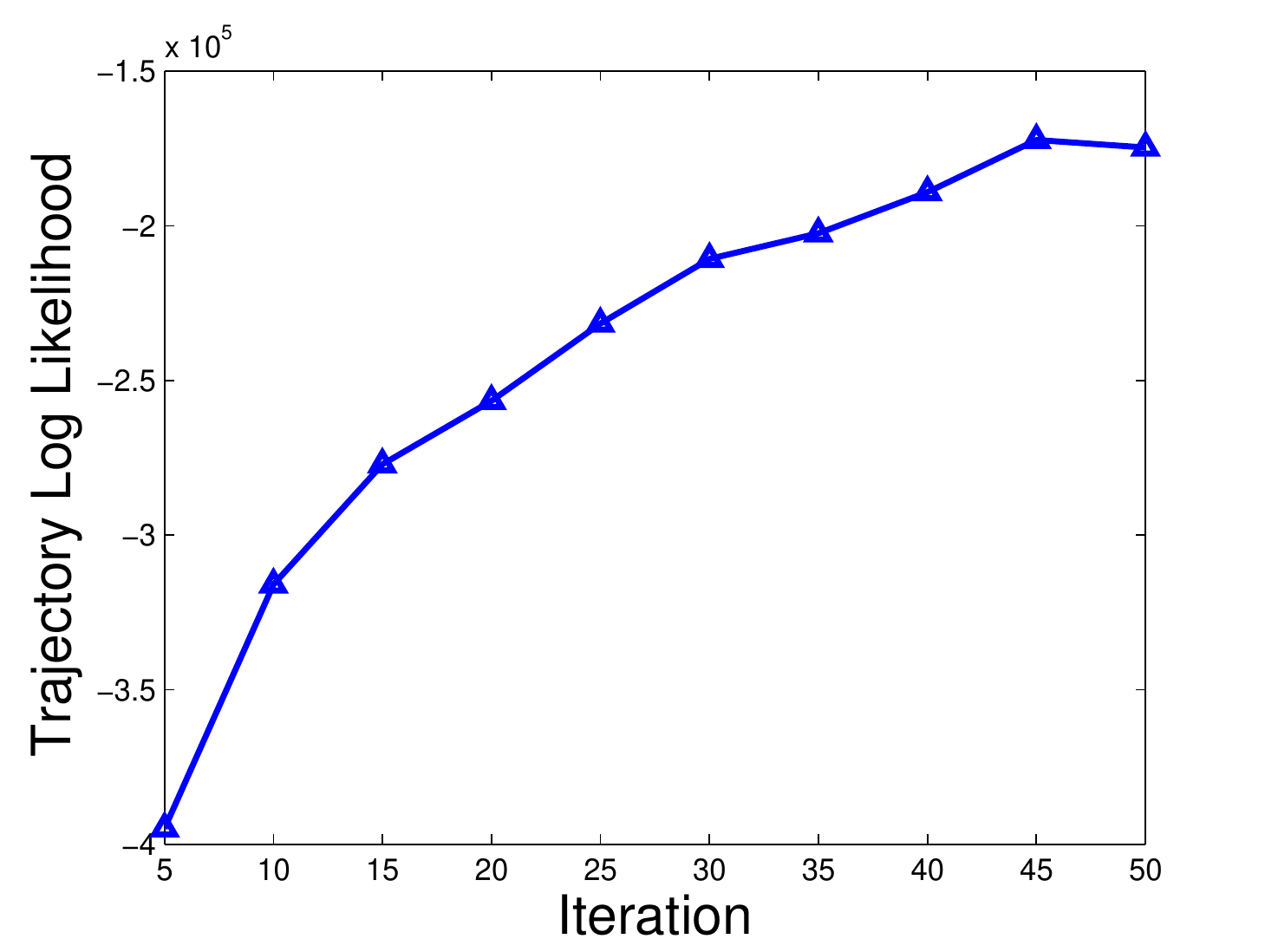}
}
\subfigure[Recall@K]{
\includegraphics[width=0.3\textwidth]{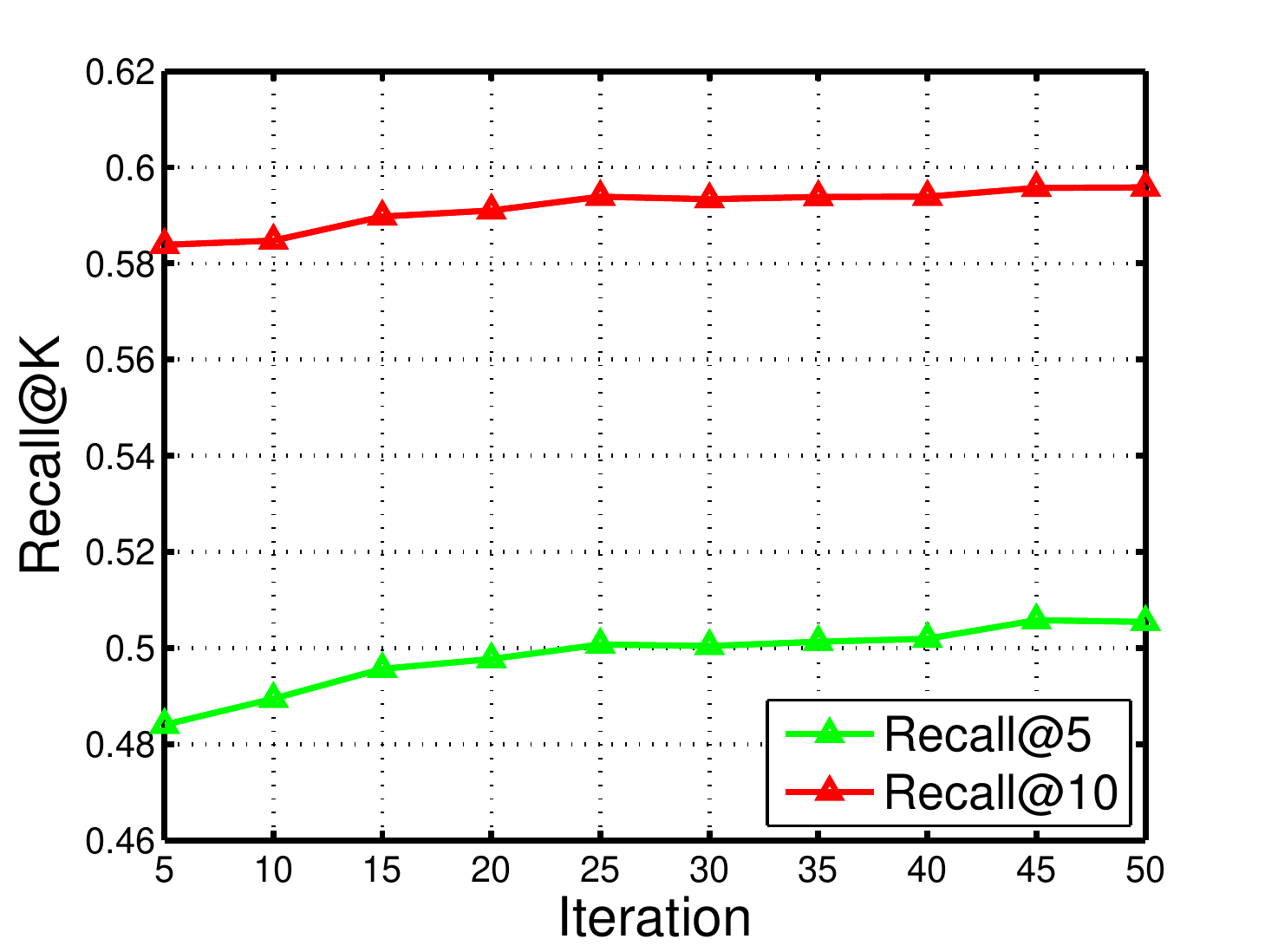}
}
\caption{Performance of the iteration number on the Brightkite dataset.} \label{fig:pti1}
\end{minipage}
\begin{minipage}{\textwidth}
\subfigure[Network log likelihood]{
\includegraphics[width=0.3\textwidth]{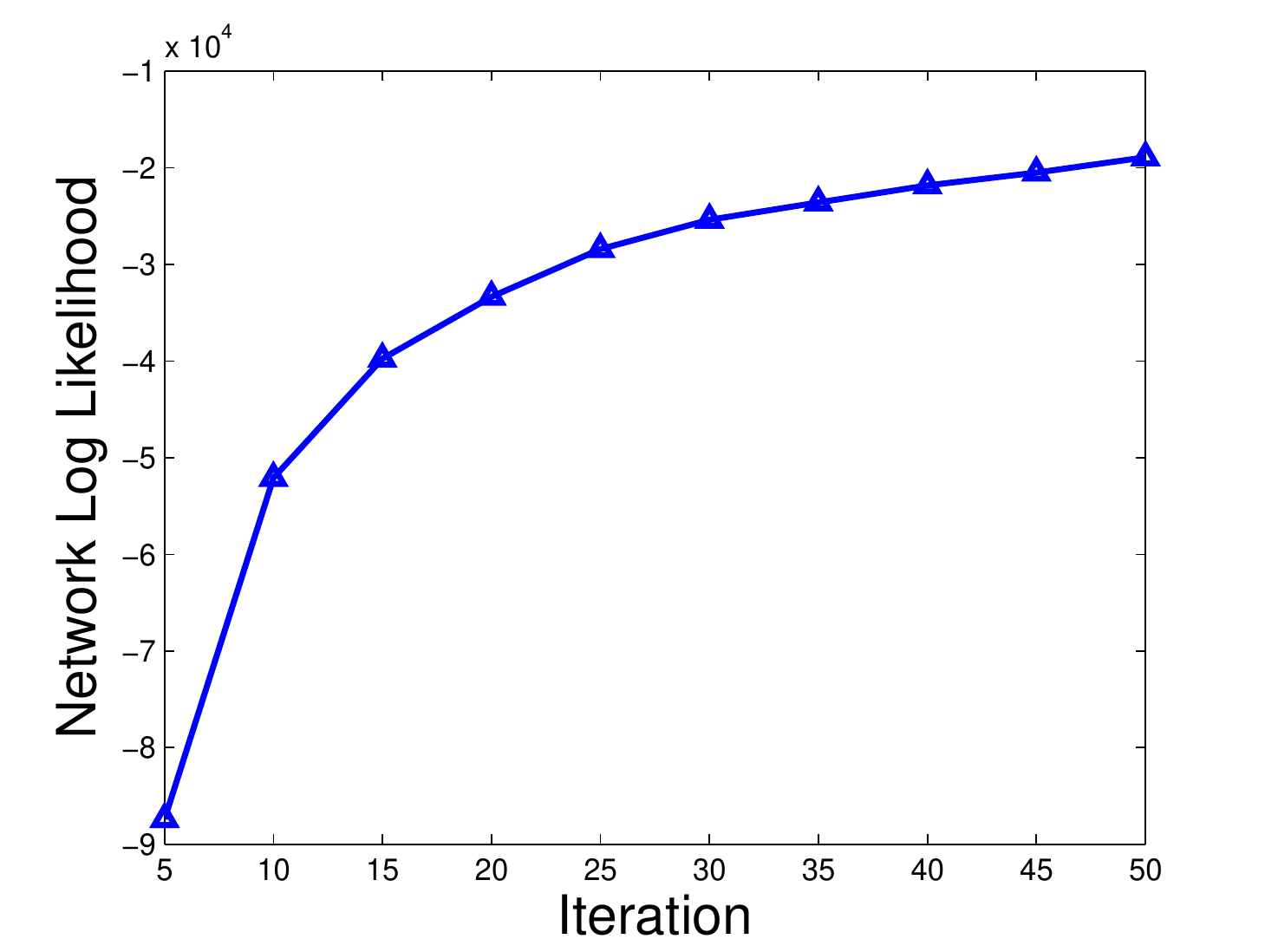}
}
\subfigure[Trajectory log likelihood]{
\includegraphics[width=0.3\textwidth]{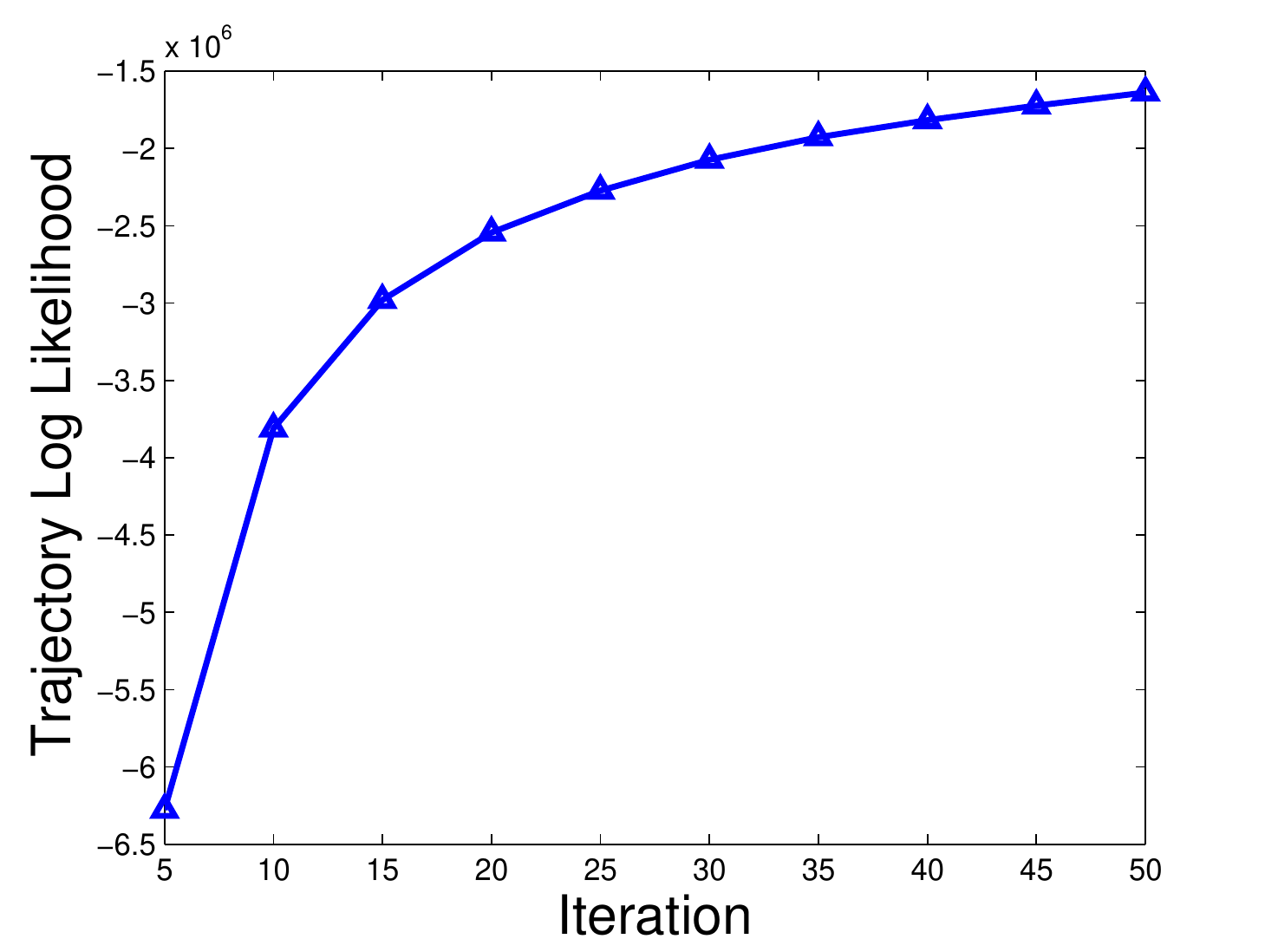}
}
\subfigure[Recall@K]{
\includegraphics[width=0.3\textwidth]{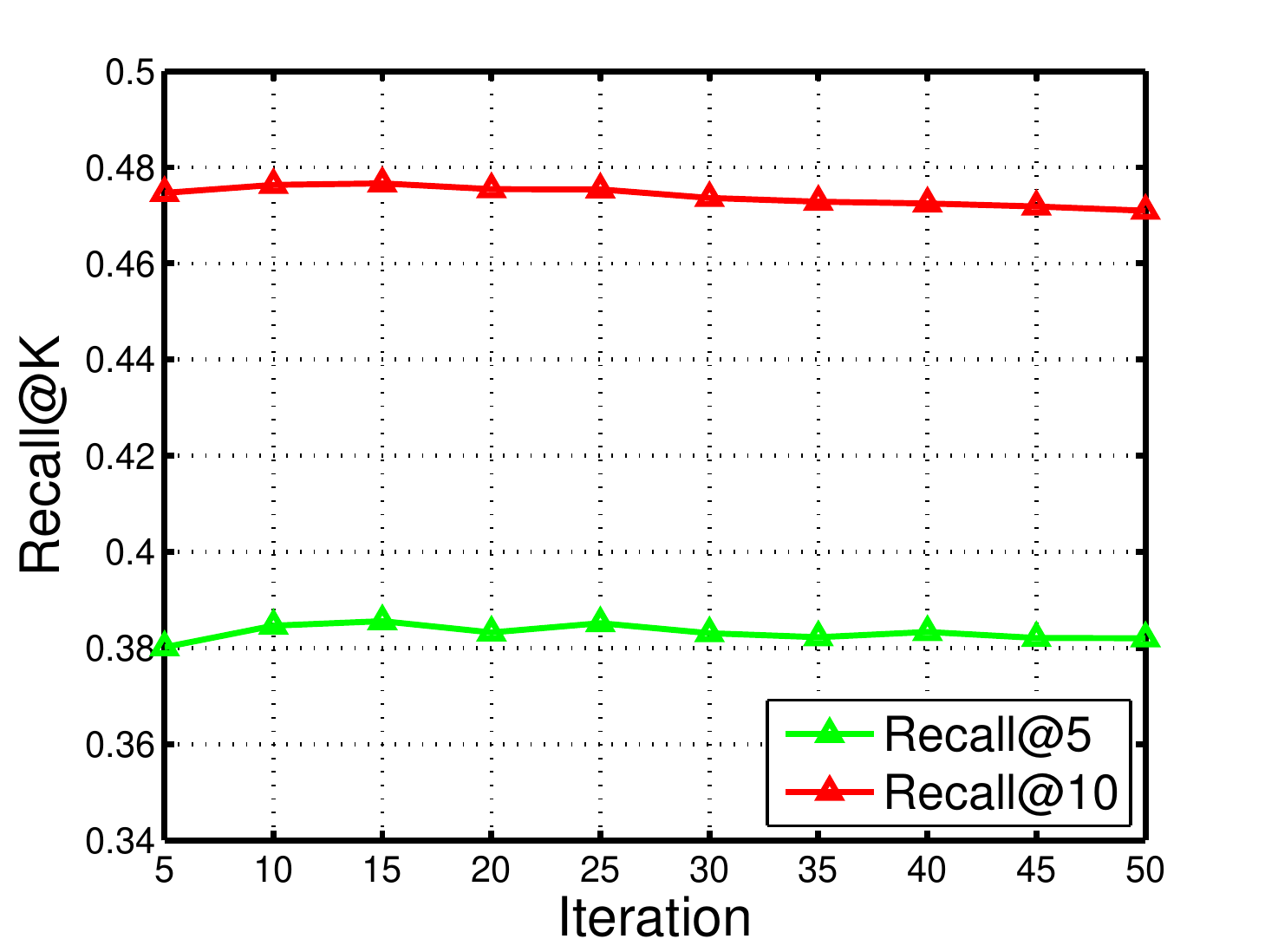}
}
\end{minipage}
\caption{Performance of the iteration number on the on the Gowalla dataset.} \label{fig:pti2}
\end{figure*}

We conduct the tuning experiments on the training sets by varying the number of iterations from $5$ to $50$. We report
the log likelihood for the network and trajectory data on the training sets and Recall@5 and Recall@10 of next location recommendation on validation sets.

Fig. \ref{fig:pti1} and \ref{fig:pti2} show the tuning results of the iteration number on both datasets. From the results we can see that our algorithm can converge within $50$ iterations on both datasets, and the growth of log likelihood slows down after 30 iterations. On the other hand, the performance of next location recommendation on validation sets is relatively stable: JNTM can yield a relatively good performance after $5$ iterations. The recall values increase slowly and reach the highest score at $45$ iteration on Brightkite and $15$ iteration on Gowalla dataset. Here Gowalla dataset converges more quickly and smoothly than Brightkite. It is mainly  because Gowalla dataset contains $3$ times more check-in data than that of Brightkite and has more enough training data. However the model may overfit before it gets the highest recall for next-location recommendation because the recall scores are not always monotonically increasing. As another evidence, the performance on new location prediction begins to drop after about $10$ iterations. To avoid the overfitting problem, we reach a compromise and find that an iteration number of $15$ and $10$ is a reasonably good choice to give good performance on Brightkite and Gowalla, respectively.

\begin{figure*}[ht]
\centering
\begin{minipage}{\textwidth}
\subfigure[Brightkite]{
\includegraphics[width=0.47\textwidth]{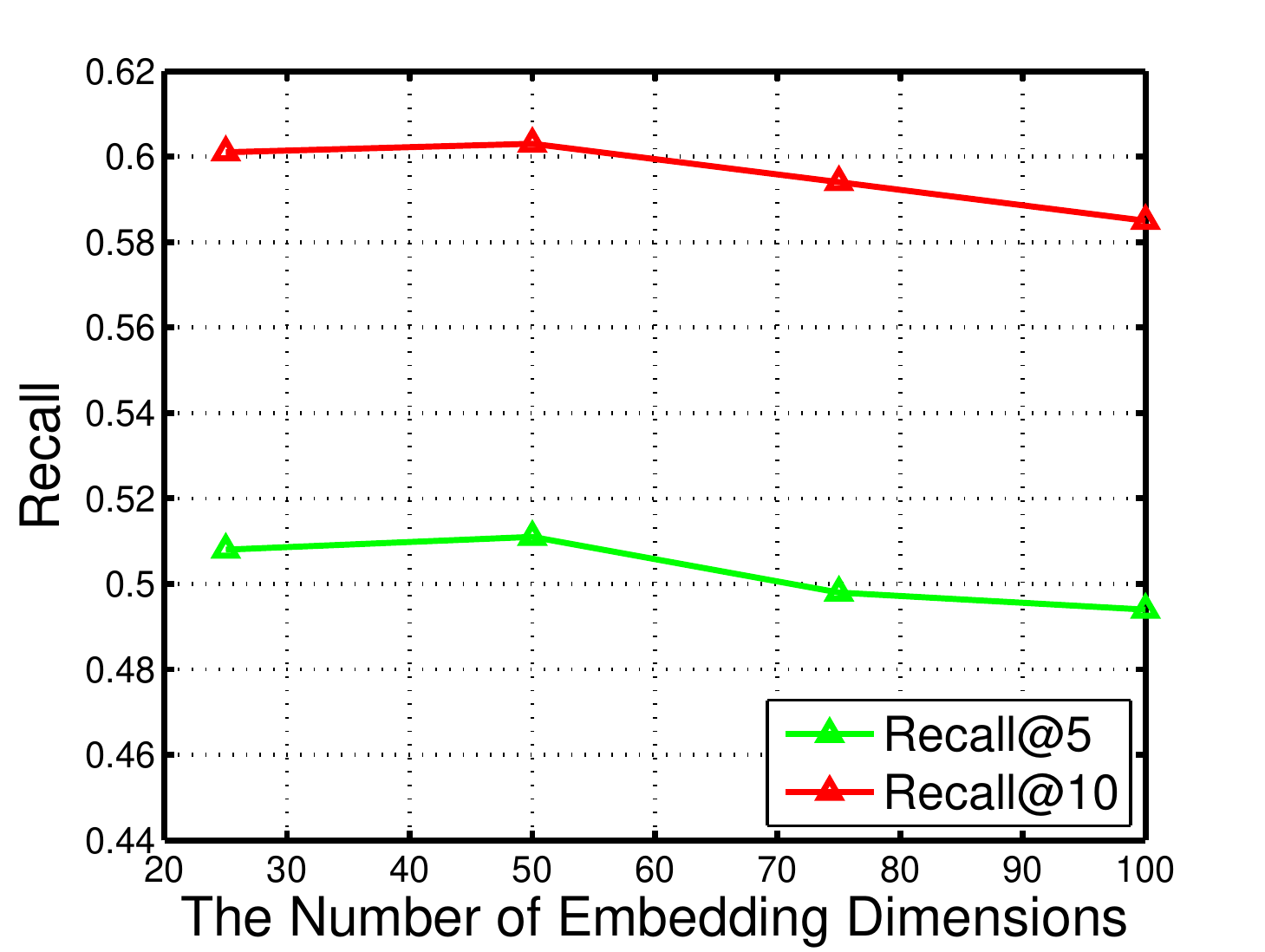}
}
\subfigure[Gowalla]{
\includegraphics[width=0.47\textwidth]{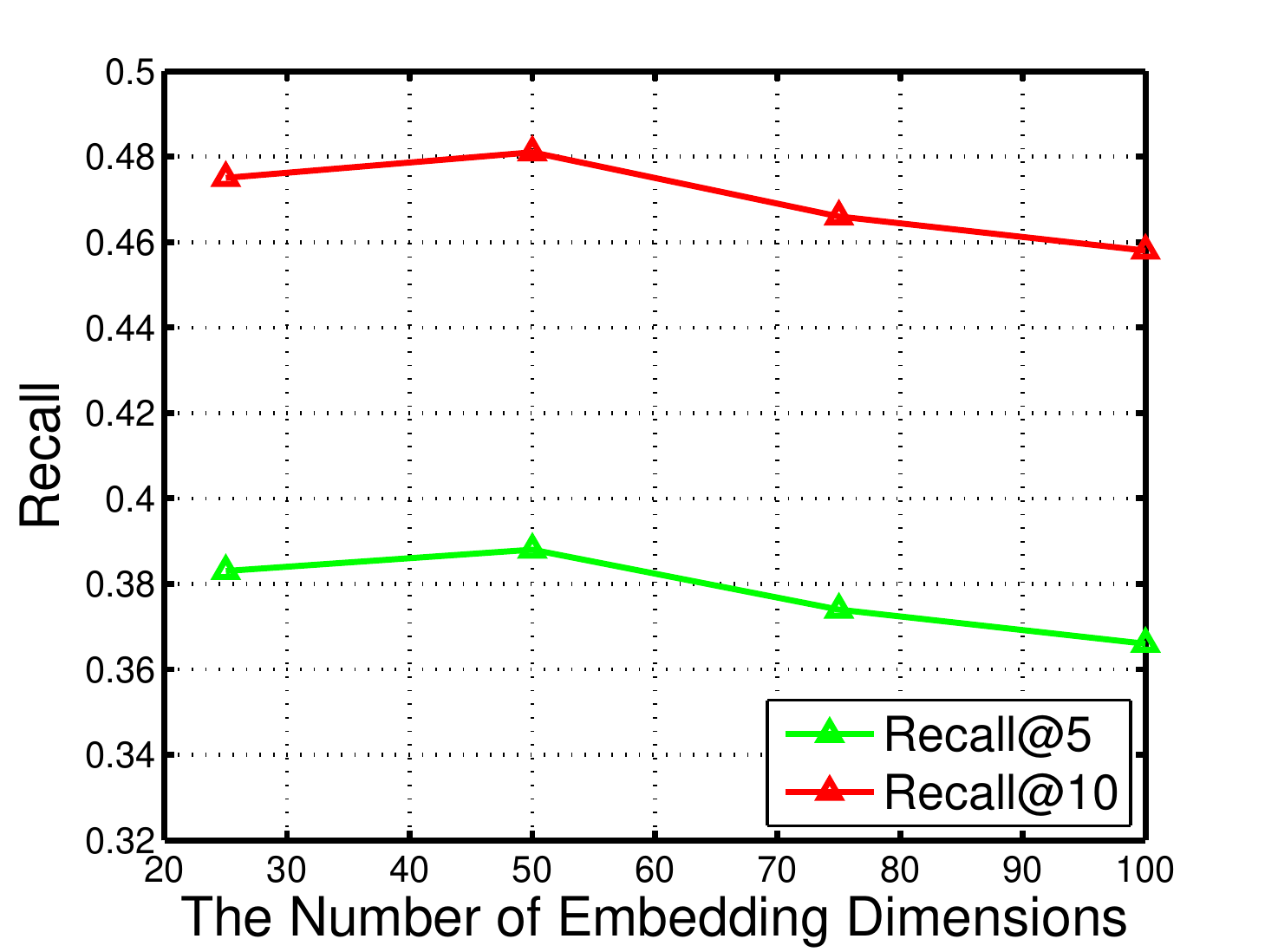}
}
\caption{Performance tuning with different dimension numbers. } \label{dimension}
\end{minipage}
\end{figure*}

The number of embedding dimensions is also vital to the performance of our model. A large dimension number will have a strong expressive ability but will also probably lead to overfitting. We conduct experiments with different embedding dimension numbers on next location recommendation and measure their performance on validation sets.
In Fig. \ref{dimension}, we can see that the performance of our algorithm is relatively stable when we vary the dimension number from $25$ to $100$. The recall values start to decrease when the  dimension number exceeds $50$. We finally set the dimension number to $50$.

\subsection{Scalability}
In this part, we conduct experiments on scalability and examine the time and space costs of our model. We perform the experiments on Gowalla dataset, and select the baseline method PRME~\cite{feng2015personalized} for comparison. We report the memory usage of both methods for space complexity analysis and running time on a single CPU for time complexity analysis. Since both methods have the same iteration number for convergence, we report the average running time per iteration for each algorithm. Running time for training and testing  is presented separately.
A major merit of neural network models is that they can be largely accelerated by supported hardware (\eg GPU). Therefore we also report the running time of a variation of our model using the GPU acceleration. Specifically, we use a single Tesla K40 GPU for training our model in the \textsc{TensorFlow}\footnote{https://www.tensorflow.org} software library. The experimental results are shown in Table~\ref{tab:scalability}.

\begin{table}[htb]
\centering
\tbl{Experimental results on memory usage (MB) and running time (minute)}{
\begin{tabular}{ccccc}
\hline
 Model & $Memory$ & $Training$ & $Testing$ & $Training\:(GPU)$\\
\hline
\hline
PRME & 550 & 2 & 52 & -\\
JNTM & 1,125 & 107 & 82 & 9\\
\hline
\end{tabular}}
\label{tab:scalability}
\end{table}

From Table~\ref{tab:scalability}, we can see that our memory usage is almost twice as much as PRME. This is mainly because we set two representations for each user (\ie friendship and preference representations), while PRME only has a preference representation. The time complexity of JNTM is $O(d^2|D|+d|V|)$, while  the time complexity of PRME is $O(d|D|)$,  where $d$ is the embedding dimensionality ($d=50$ in our experiments). Hence the running time of JNTM is about $d$ times as much as that of PRME. Although our model has a longer training time than PRME, the time cost of JNTM for testing is almost equivalent to that of PRME. On average, JNTM takes less than $30ms$ for a single location prediction, which is efficient to provide online recommendation service after the training stage. Moreover, the GPU acceleration offers $12$x speedup for the training process, which demonstrates that our model JNTM can be efficiently learned with supported hardware.

\section{Conclusion and Future Work}
In this paper, we presented a novel neural network model by jointly model both social networks and mobile trajectories.
In specific, our model consisted of two components: the construction of social networks and the generation of mobile trajectories. We first adopted a network embedding method for the construction of social networks. We considered four factors that influence the generation process of mobile trajectories, namely user visit preference, influence of friends, short-term sequential contexts and long-term sequential contexts. To characterize the last two contexts, we employed the RNN and GRU models to capture the sequential relatedness in mobile trajectories at different levels, i.e., short term or long term. Finally, the two components were tied by sharing the user network representations. On two important application tasks, our model was consistently better than several competitive baseline methods.
In our approach, network structure and trajectory information complemented each other. Hence, the improvement over baselines was more significant when either network structure or trajectory data is sparse.

Currently, our model does not consider the GPS information, \ie a check-in record is usually attached with a pair of longitude and latitude values. Our current focus mainly lies in how to jointly model social networks and mobile trajectories. As the future work, we will study how to incorporate the GPS information into the neural network models. In addition, the check-in location can be also attached with categorical labels. We will also investigate how to leverage  these semantic information to improve  the performance. Such semantic information can be utilized for the explanation of the generated recommendation results. Our current model has provides a flexible neural network framework to characterize LBSN data. We believe  it will inspire more follow-up studies along this direction.


\begin{acks}
The authors thank the anonymous reviewers for their valuable and constructive comments.
\end{acks}

\bibliographystyle{ACM-Reference-Format-Journals}
\bibliography{main}

\received{}{}{}


\medskip

\end{document}